%% file: main.tex
\documentclass[review=false,screen=false,acmsmall,anonymous=false]{acmart}
\AtBeginDocument{%
  }

\setcopyright{acmcopyright}
\copyrightyear{2023}
\acmYear{2023}
\acmDOI{XXXXXXX.XXXXXXX}

\usepackage{xspace}
\usepackage{tabularx}
\usepackage{tcolorbox}
\usepackage{hyperref}
\usepackage{blindtext}
\usepackage{caption}
\usepackage{subcaption}
\usepackage{float}
\usepackage{paralist}
\usepackage{needspace}
\usepackage{pifont}
\usepackage{multirow}
\usepackage{multicol}
\usepackage{url}
\usepackage{mdframed}
\usepackage{arydshln}
\usepackage{tabularray}
\UseTblrLibrary{booktabs}
 \UseTblrLibrary{siunitx}
\usepackage{needspace}
\usepackage{pgfplots}
\pgfplotsset{compat=1.18}
\usepgfplotslibrary{statistics}
\usepackage{tikz}
\usetikzlibrary{shapes.geometric, arrows, positioning}

\newcommand{\circled}[1]{\tikz[baseline=(char.base)]{
            \node[shape=circle,draw, fill=black, minimum size=8pt, inner sep=0.1pt, outer sep=0pt] (char) {\scriptsize \textcolor{white}{#1}};}}

\newcommand{\participantscount}[0]{50 }
\newcommand{\framework}[0]{\textsc{SDC-Alabaster}\xspace}
\newcommand{\frameworklong}[0]{SDC hum\textbf{A}n-in-the \textbf{L}oop simul\textbf{A}tion-\textbf{BAS}ed \textbf{T}esting s\textbf{E}lf-driving ca\textbf{R}s\xspace}

\newcommand{\RQ}[1]{\texorpdfstring{RQ\textsubscript{#1}}{RQ#1}}
\newcommand{\beamng}[0]{BeamNG.tech\xspace}
\newcommand{\carla}[0]{CARLA\xspace}

\newcommand{\vorpx}[0]{\textsc{vorpX}\xspace}

\newcommand{\sdcscissor}[0]{\textsc{SDC-Scissor}\xspace}
\newcommand{\vive}[0]{HTC Vive Pro 2\xspace}

\newcommand{\mannwhitneyu}[0]{Wilcoxon rank-sum\xspace}
\newcommand{\shapirowilk}[0]{Shapiro-Wilk\xspace}
\newcommand{\varghadelaney}[0]{Vargha-Delaney\xspace}
\newcommand{\veryunsafe}[0]{very unsafe\xspace}
\newcommand{\unsafe}[0]{unsafe\xspace}
\newcommand{\neutral}[0]{neutral\xspace}
\newcommand{\safe}[0]{safe\xspace}
\newcommand{\verysafe}[0]{very safe\xspace}

\newcommand{\eg}[0]{e.g.,\xspace}
\newcommand{\rqO}{When and why do safety metrics of simulation-based test cases of self-driving cars match human perception?\xspace}
\newcommand{\rqI}{To what extent does the OOB safety metric for simulation-based test cases of SDCs align with human safety assessment?\xspace}
\newcommand{\rqII}{To what extent does the safety assessment of simulation-based SDC test cases vary when humans can interact with the SDC?\xspace}
\newcommand{\rqIII}{
What are the main reality-gap characteristics perceived by humans in SDC test cases?
\xspace}

\mdfsetup{backgroundcolor=gray!10}
\newcounter{findingcounter}
\newenvironment{finding}
{
\vspace{5pt}
\Needspace{8\baselineskip}
\begin{mdframed}
\refstepcounter{findingcounter}
\emph{Finding \thefindingcounter:}
}
{ 
\end{mdframed}
\vspace{5pt}
}

\begin{document}
\title{How does Simulation-based Testing for Self-driving Cars match Human Perception?}
\input{sections/00-authors}
\input{sections/00-abstract}

\maketitle

\input{sections/01-introduction}
\input{sections/02-1-background}
\input{sections/02-2-methodology}
\input{sections/03-results}
\input{sections/04-discussion}

\input{sections/05-threats-to-validity}
\input{sections/06-related-work}

\input{sections/07-conclusion}

\bibliographystyle{ACM-Reference-Format}
\bibliography{extracted}
\clearpage
\input{sections/09-appendices}

\end{document}

%% file: sections/00-authors.tex
\author{Christian Birchler}
\email{birc@zhaw.ch}
\email{christian.birchler@unibe.ch}
\affiliation{%
 \institution{Zurich University of Applied Sciences \& University of Bern}
 \country{Switzerland}}


\author{Tanzil Kombarabettu Mohammed}
\email{tanzil.kombarabettumohammed@uzh.ch}
\affiliation{%
  \institution{University of Zurich}
  \city{Zurich}
  \country{Switzerland}
}

\author{Pooja Rani}
\email{rani@ifi.uzh.ch}
\affiliation{%
  \institution{University of Zurich}
  \city{Zurich}
  \country{Switzerland}
}
\author{Teodora Nechita}
\email{neci@zhaw.ch}
\affiliation{%
  \institution{Zurich University of Applied Sciences}
  \city{Winterthur}
  \country{Switzerland}}

\author{Timo Kehrer}
\email{timo.kehrer@unibe.ch}
\affiliation{%
 \institution{University of Bern}
 \city{Bern}
 \country{Switzerland}}

\author{Sebastiano Panichella}
\email{panc@zhaw.ch}
\affiliation{%
  \institution{Zurich University of Applied Sciences}
  \city{Winterthur}
  \country{Switzerland}}

\renewcommand{\shortauthors}{Birchler et al.}

%% file: sections/00-abstract.tex
\begin{abstract}
Software metrics such as coverage and mutation scores have been extensively explored for the automated quality assessment of test suites.
While traditional tools rely on such quantifiable software metrics, the field of self-driving cars (SDCs) has primarily focused on simulation-based test case generation using quality metrics such as the out-of-bound (OOB) parameter to determine if a test case fails or passes.
However, it remains unclear to what extent this quality metric aligns with the human perception of the safety and realism of SDCs, which are critical aspects in assessing SDC behavior.
To address this gap, we conducted an empirical study involving 50 participants to investigate the factors that determine how humans perceive SDC test cases as safe, unsafe, realistic, or unrealistic.
To this aim, we developed a framework leveraging virtual reality (VR) technologies, called \framework, to immerse the study participants into the virtual environment of SDC simulators.
Our findings indicate that the human assessment of the safety and realism of failing and passing test cases can vary based on different factors, such as the test's complexity and the possibility of interacting with the SDC.
Especially for the assessment of realism, the participants' age as a confounding factor leads to a different perception.
This study highlights the need for more research on SDC simulation testing quality metrics and the importance of human perception in evaluating SDC behavior.
\end{abstract}

%% file: sections/01-introduction.tex
\section{Introduction}\label{sec:introduction}
In recent years, the development of autonomous systems has impacted our society in many aspects of our life~\cite{AraujoMV23,BohnePK23}.
For instance, humans no longer rely on vacuuming their houses or mowing their grasses manually; nowadays, we have robots that do (and will do) much of our chores~\cite{BBC-robots}.
However, specific safety-critical instances of such autonomous systems such as unmanned aerial vehicles (UAVs) and self-driving cars (SDCs)~\cite{10.1145/3377811.3380353,DiSorboTOSEM2023,surrealist,DBLP:journals/ethicsit/Stilgoe21,DBLP:conf/icst/Jahangirova0T21} may experience failures that
can harm humans or damage the environment~\cite{TheGuardian-2018}.

\begin{figure}[t]
    \centering
    \begin{subfigure}{0.45\linewidth}
        \centering
        \frame{\includegraphics[width=0.7\textwidth]{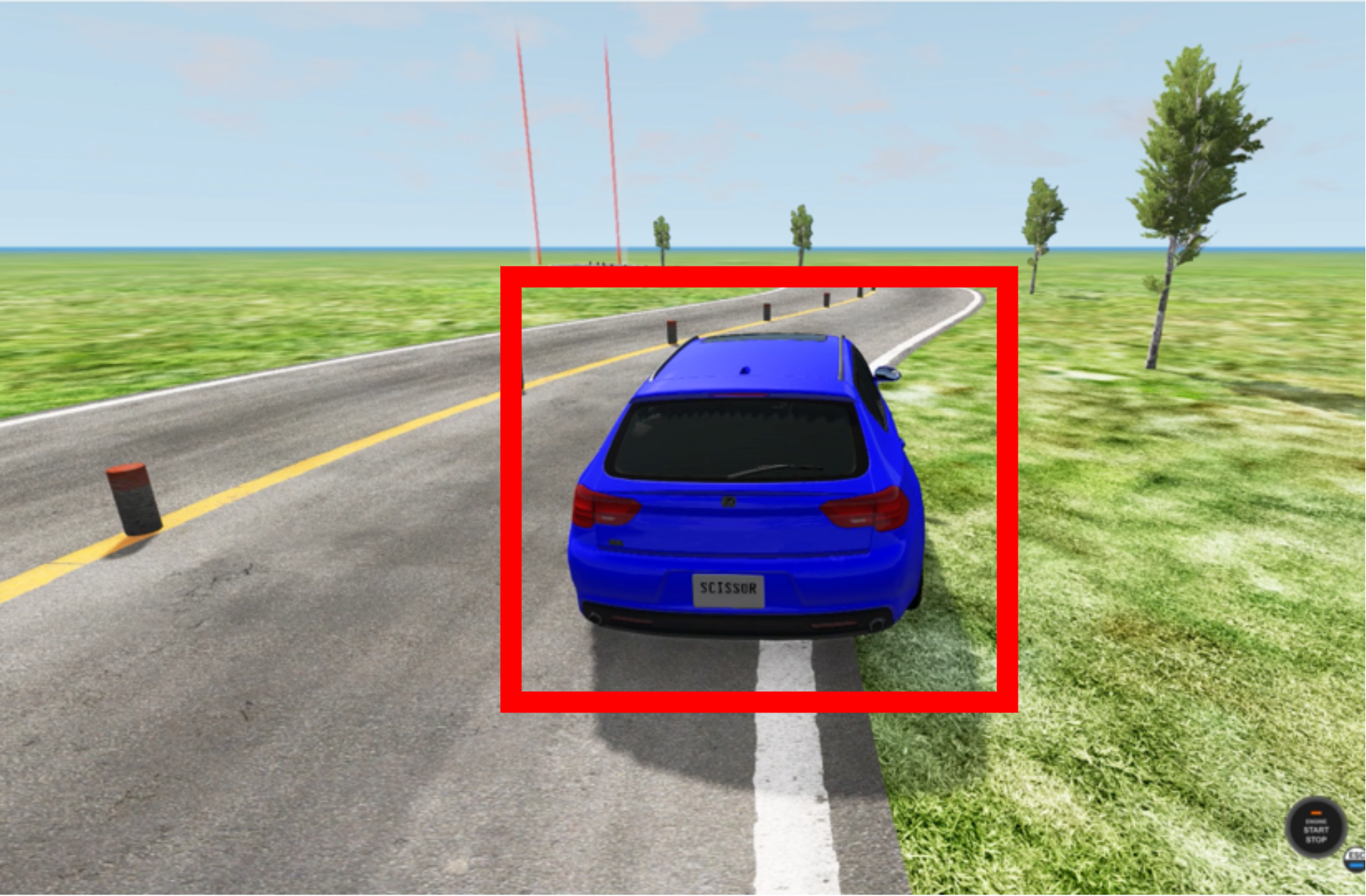}}
        \caption{Failing Test: SDC driving off-lane (unsafe).}
        \label{fig:03-invalid-oob}
    \end{subfigure}
    \hfill
    \begin{subfigure}{0.45\linewidth}
        \centering
        \frame{\includegraphics[width=0.7\textwidth]{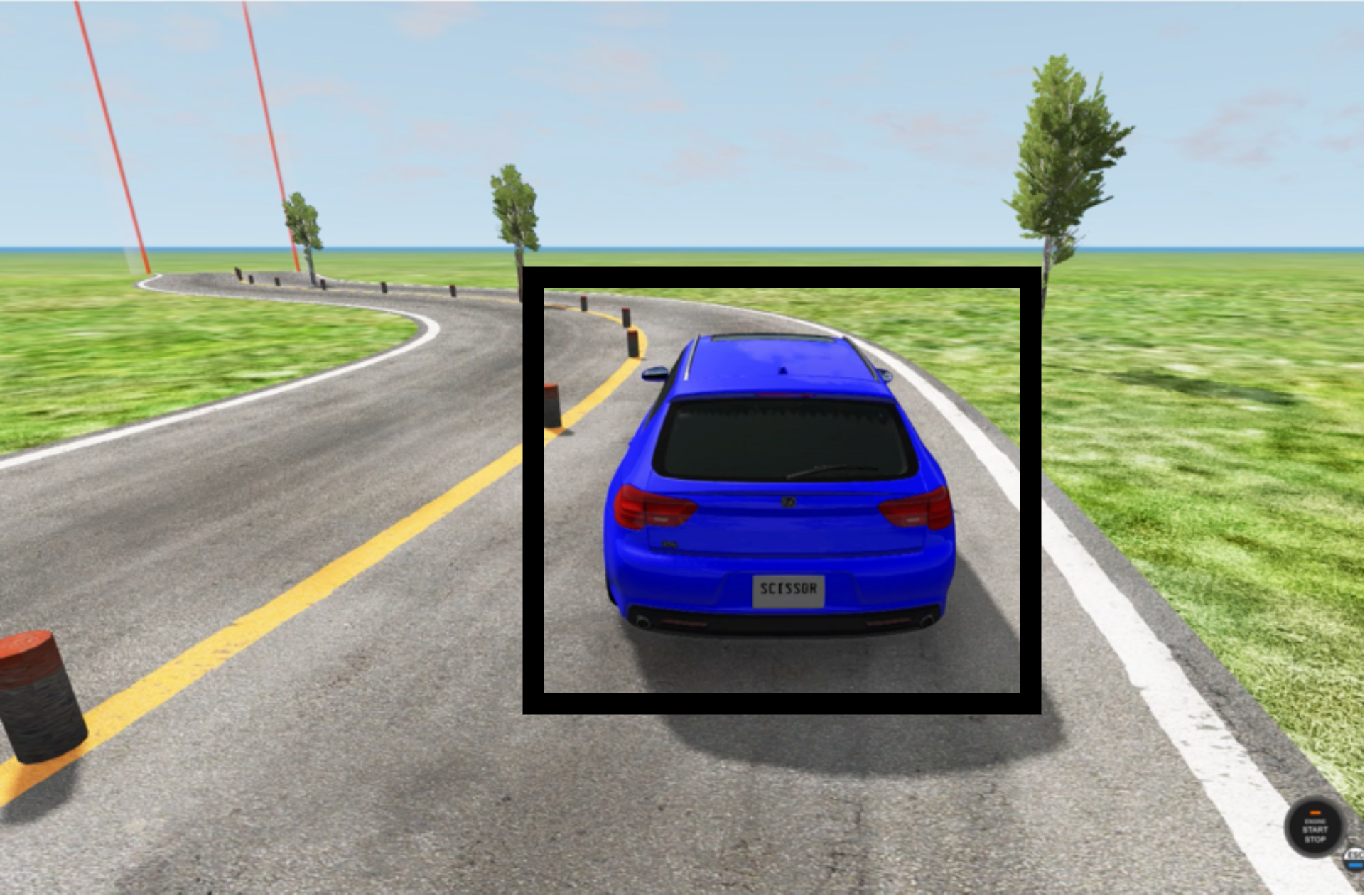}}
        \caption{Passing Test: SDC driving in-lane (safe) .}
        \label{fig:04-valid-oob}
    \end{subfigure}
            \vspace{-3mm}
    \caption{
    Examples of simulation-based tests of an SDC.}
    \label{fig:oob-metric}
        \vspace{-5mm}
\end{figure}

Testing safety-critical autonomous systems is crucial to avoid harmful incidents in real environments~\cite{DBLP:conf/ijcai/Wotawa19,1438843,DBLP:conf/icse/GambiHF19,AbdessalemPNBS20,DBLP:journals/ese/BirchlerKBGP23}.
To that end, simulation environments have been widely adopted to test cyber-physical systems (CPS) in general \cite{beamNG,DosovitskiyRCLK17,nvidia_drive}, and SDCs in particular~\cite{DosovitskiyRCLK17,beamNG}. 
As opposed to real-world testing, simulation-based testing is easier to replicate, is more cost-efficient, and can be as effective as field testing~\cite{HildebrandtE21,DosovitskiyRCLK17}.
Figure~\ref{fig:oob-metric} illustrates two test cases where an SDC model is deployed in a virtual environment, and the simulated car is expected to behave according to the control algorithms. 
A test case is said to pass if the car's behavior can be considered safe, while unsafe behavior constitutes a failing test case.
Figure~\ref{fig:03-invalid-oob} shows an unsafe behavior (failing test) as the SDC drives off the lane, while Figure~\ref{fig:04-valid-oob} shows a passing test.

Current research on simulation-based test case generation (STSG) of SDCs relies on an oracle that determines if a system under test is safe or unsafe based 
on a limited set of safety metrics~\cite{DBLP:conf/sbst/PanichellaGZR21,GambiJRZ22,DBLP:journals/ese/BirchlerKBGP23},  
particularly the out-of-bound (OOB) metric. 
The metric is largely adopted for assessing the safety behavior in STSG~\cite{GambiJRZ22,DBLP:conf/sbst/PanichellaGZR21,NguyenHG21}.
Both test cases illustrated in Figure~\ref{fig:oob-metric} are classified using the OOB metric~\cite{DBLP:journals/tosem/BirchlerKDPP23} and align with the human perception of safety.

\begin{figure}[t]
    \centering
    \begin{subfigure}{0.45\textwidth}
        \centering
        \frame{\includegraphics[width=0.7\textwidth]{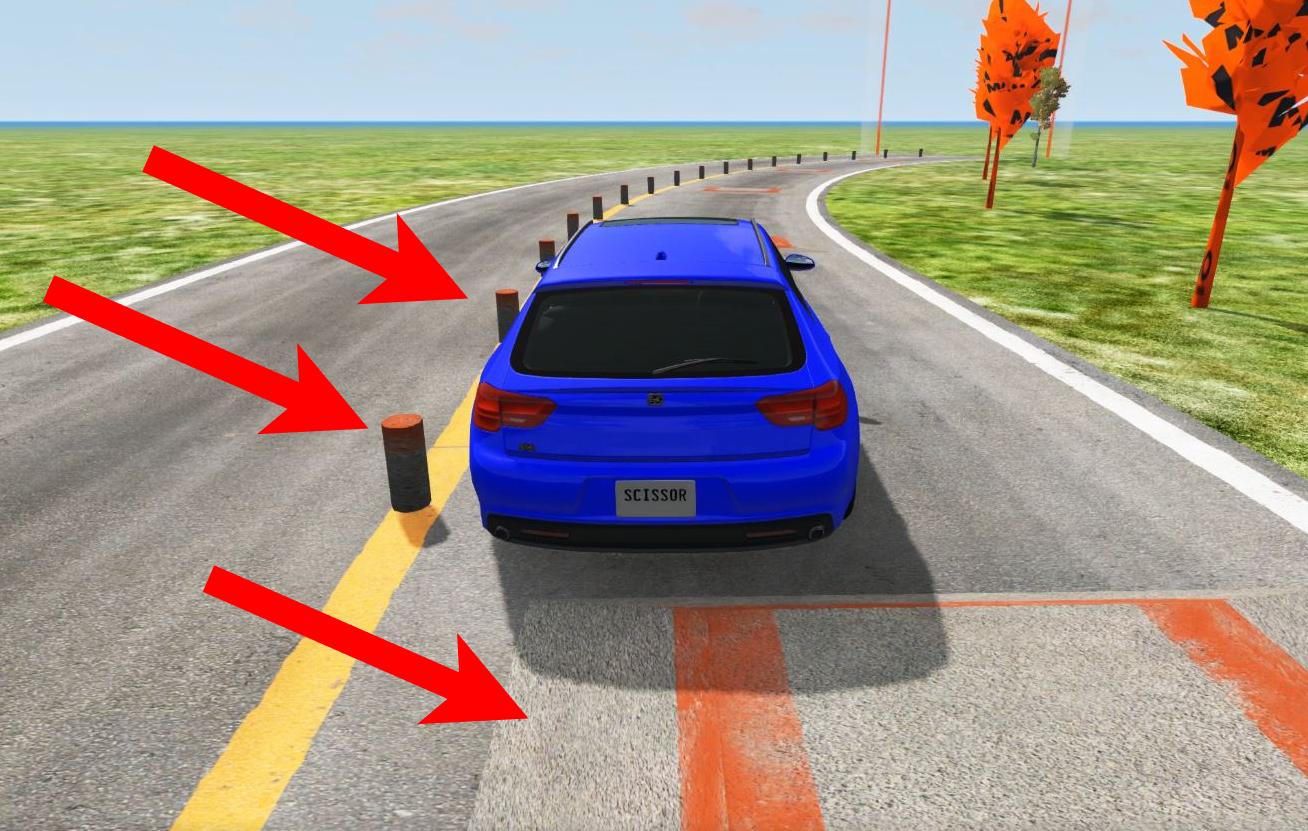}}
        \caption{SDC in \beamng driving with 50 km/h close to obstacles}
        \label{fig:01-motivating-example}
    \end{subfigure}
    \hfill
    \begin{subfigure}{0.45\textwidth}
        \centering
        \frame{\includegraphics[width=0.7\textwidth]{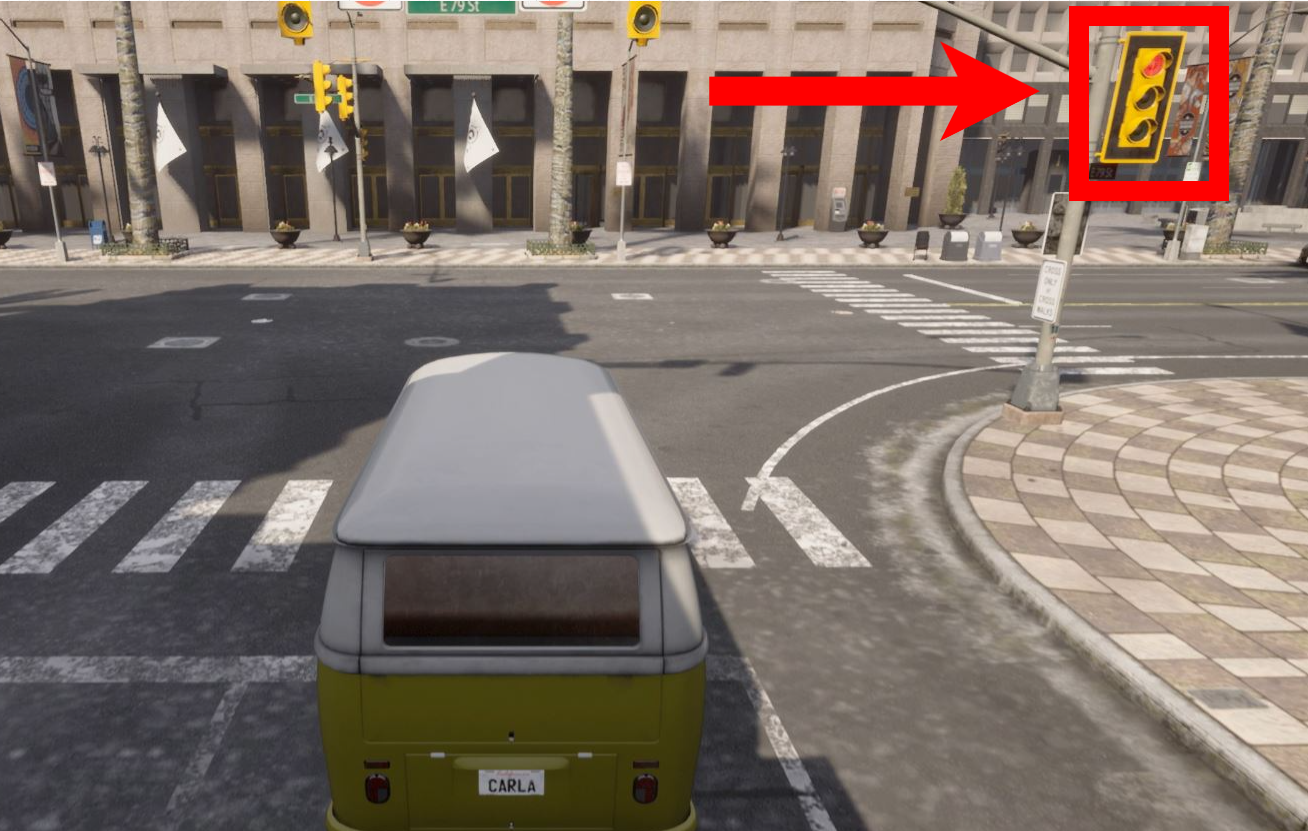}}
        \caption{SDC in \carla crossing a red signal without stopping}
        \label{fig:02-motivating-example}
    \end{subfigure}
    \caption{Examples of unsafe tests with valid OOB criteria}
    \label{fig:motivating-example}
\end{figure}

However, it is yet unclear whether STSG metrics (e.g., OOB) serve as meaningful oracles for assessing the safety 
behavior of SDCs.
For instance, the test cases in Figure~\ref{fig:motivating-example} are marked pass
according to the OOB metric, as the SDC is 
keeping the lane. 
On the contrary, 
from a human standpoint, we can consider the behavior of the SDC hardly as safe.
%
In the first test case using the \beamng simulator~\cite{beamng-tech-website}, as shown in Figure~\ref{fig:01-motivating-example}, the SDC approaches solid delineators after ignoring a speed bump. Despite maintaining its lane at a speed of 50 km/h, there is a high risk of an accident in classifying this test case as a technical pass based on the OOB metric.
%
In the second test case using the \carla simulator~\cite{DosovitskiyRCLK17}, shown in Figure~\ref{fig:02-motivating-example}, the SDC ignores the red signal. Since the car stays in the lane, it meets the OOB metric, leading to a false passing test case.

Inspecting the OOB metric reveals that it is measured at a single point in time in simulation, which is insufficient to identify unsafe behaviors. 
For instance, Figure~\ref{fig:01-motivating-example} shows the speed bumps on the right lane, and evaluating the SDC at a single point is insufficient to assess its safety over these speed bumps.
In such cases, having a time window will be more informative to assess the overall SDC behavior.
Unlike real-world speed bumps, which are smooth and rounded, the test bumps have sharp edges that damage the SDC even at reasonable speeds (from a human viewpoint).
Similarly, Figure~\ref{fig:02-motivating-example} shows another instance where we observe the red light signal, but the SDC ignores it. It is unclear whether the red signal was already there before the SDC drove past it or the signal turned red just after the SDC analyzed the simulation scene. 
We hypothesize that current simulation-based testing of SDCs does not always align with the human perception of safety ~\cite{GambiJRZ22,DBLP:conf/sbst/PanichellaGZR21,NguyenHG21} and realism ~\cite{wang2021exploratory,afzal2021simulation,NgoBR21,RewayHWHKR20}, which are 
relevant aspects impacting the effective
assessment of simulated-based test cases.
Hence, our primary goal is to understand and characterize this mismatch by answering the following research question:
\begin{tcolorbox}
When and why do safety metrics of simulation-based test cases of SDCs match human perception?
\end{tcolorbox}


To answer our general research question 
(i.e., addressing the problem of \textit{safety} and \textit{realism} of test cases that are described in our motivating examples), we conducted an empirical study involving 50 participants using our framework named \framework.
The framework employs virtual reality (VR) technologies~\cite{DBLP:conf/hri/SilveraBA22} (i) to immerse humans in virtual SDCs so that they can sense and experience the virtual environment as similar as possible to the real world, and (ii) to enable SDC developers and researchers to analyze the human perception of \textit{safety} and \textit{realism} of SDC test cases.
The participants in our study are asked to assess the level of \textit{safety} and \textit{realism} of multiple, diverse simulation-based test cases.
Moreover, we provide the participants to experience simulation-based test cases in which they have the possibility to influence the behavior of (i.e., interact with) the SDC. 
For this purpose, we experimented with two representative SDC simulators as virtual environments, \beamng and \carla, which are widely used in academia and industry~\cite{DBLP:conf/itsc/2022,GambiJRZ22}. 


The paper contributes and complements previous research as follows:
\begin{itemize}
    \item we propose the \framework framework to assess simulation-based SDC test cases from a human point of view with VR;
    \item we investigate the perceived level of \textit{safety} and \textit{realism} of simulation-based SDC test cases by conducting an empirical study with 50 participants. We publicly share a replication package with the code to reproduce our results (Section~\ref{sec:data-availability});
    \item we develop a taxonomy on impacting factors on the perceived realism of SDC simulators and provide a discussion on confounding factors and implications of our work.
\end{itemize}

The paper covers background (Section~\ref{sec:background}), study design (Section~\ref{sec:study-design}), our framework,  
experiments, and methodology. Section~\ref{sec:results} presents our results, followed by discussions in Section~\ref{sec:discussion} and threats to validity in Section~\ref{sec:threats-to-validity}. We discuss related work and conclusions in Section~\ref{sec:related-work} and  Section~\ref{sec:conclusion}.


%% file: sections/02-1-background.tex
\section{Background}\label{sec:background}

This section provides a background on existing technologies used in our study, such as simulators, test generators, and test runners for SDCs, as well as VR technology.

\subsection{SDC simulators}
We investigate when the safety metrics of STSG for SDCs match the human perception.
To answer this question, we use two state-of-the-art SDC simulators namely \beamng, and \carla.
They are among the used SDC simulators widely used in academia and practice~\cite{GambiJRZ22,DBLP:conf/aaai/NairSAK21,10.1007/978-3-030-60508-7_9,DBLP:journals/sensors/Gutierrez-Moreno22,DBLP:conf/sbst/PanichellaGZR21,DosovitskiyRCLK17}.
Furthermore, they implement fundamentally different physics behaviors.

\subsubsection{\beamng}
We use \beamng simulator as a well-known reference technology used in recent years in several studies and software engineering competitions on testing SDCs~\cite{BeamNGresearch,DBLP:conf/sbst/PanichellaGZR21,DBLP:journals/ese/BirchlerKBGP23,DBLP:journals/tosem/BirchlerKDPP23,GambiJRZ22}.
The \beamng simulator comes along with a soft-body physics engine that allows the simulation of body deformations and therefore more realistic simulations regarding crashes and impacting forces on objects.

\subsubsection{\carla}
Another widely used simulator in academia and practice is \carla~\cite{DosovitskiyRCLK17, 10.1007/978-3-030-60508-7_9, DBLP:conf/aaai/NairSAK21, DBLP:journals/mta/HuelamoEBBGAAL22, DBLP:conf/infocom/ZhangFCJZX20, DBLP:journals/sensors/Gutierrez-Moreno22}.
The differences between \carla and \beamng are twofold.
On the one hand, \carla comes with a rigid-body physics engine, which works differently than the soft-body physics engine of \beamng.
A rigid-body simulation environment does not deform objects; e.g., when a crash happens, the objects remain rigid.

\subsection{Test generators \& Test Runner}
\label{sec:test-generator-and-runners}
Both simulators require descriptions of the test case scenarios and we use existing test generators to automatically generate test cases for them.
Concretely, we use test generators from the tool competition of the \textit{Search-Based Software Testing} workshop~\cite{DBLP:conf/sbst/PanichellaGZR21,GambiJRZ22}.
The actual road in the simulation environments is the result of interpolating the road points that are generated by the test generator.


In order to run test cases in simulation environments, we need a test runner that manages the execution of the test cases and reports the test outcomes.
We use the \sdcscissor~\cite{DBLP:journals/ese/BirchlerKBGP23} tool, which integrates a test selection strategy for simulation-based test cases.
We use \sdcscissor since it has implemented a test runner that monitors the OOB metrics, which is suitable for our study.


\subsection{Virtual reality}\label{sec:virtual-reality-background}
The notion of VR refers to the immersive experience of users being inside a virtual world.
In our study, we want to provide the study participant with an immersive experience of the test cases, to have more accurate feedback on their perception of the safety and realism of SDC.
We leverage VR headsets and tooling for the simulation environments to achieve this goal.

\subsubsection{Headset \& VR connection with simulation environments}
We use the \vive headset to provide the study participants with a 360° VR experience, which offers an unrestricted view compared to a standard monitor. The headset connects via wire to an external device with a dedicated GPU for high-resolution VR rendering.
Most SDC simulators do not support VR out of the box.
This is also the case for \beamng and \carla.
Therefore, for our study, we use third-party tools to enable the missing VR support for both simulators.

For \beamng, we use \vorpx, a specialized tool to transform any visual output to the screen to a compatible input for VR headsets so that it provides
an immersive feeling for the user.
The \vorpx software gives a broader view angle when wearing a VR headset.
The user can move the head and can explore the virtual environment according to its head movement. 
In the case of the \carla simulator, Silvera et al.~\cite{DBLP:conf/hri/SilveraBA22} implemented an extension of \carla, allowing the simulator to be compatible with the \vive VR headset.
When launching the \carla application, passing the \texttt{-VR} flag puts the simulator into VR mode so that can be used with the headset. 

%% file: sections/02-2-methodology.tex
\section{Methodology}\label{sec:study-design}


Overall, our research aims to explore how safety metrics, i.e., OOB, match human perception. 
Specifically, we investigate the factors that make simulation-based SDC test cases safe or unsafe. 
Hence, with \framework (see Section \ref{sec:approach}), we conducted an empirical study involving 50 participants (recruiting explained in Section \ref{sec:survey}), with several steps (summarized by Figure \ref{fig:experiment-procedure}) devised to collect different types of evidence and data to answer our main question: \textit{\rqO} 
For this purpose, the usage of \framework immerses the study participants in virtual SDCs within widely used virtual environments, thanks to VR technologies (as detailed in Section \ref{sec:sudy-context}). 

\subsection{Research questions}

We structured our study around three main research questions (RQs).
\subsubsection{\RQ{1}: Human-based assessment of safety}
Our first research question is:

\begin{tcolorbox}
\RQ{1}: \rqI
\end{tcolorbox}

\RQ{1} explores participants' perceptions of SDC test failures and safety levels with and without VR technology.
We hypothesize that the OOB safety metric in software engineering may not align with human safety perception.
We evaluate alignment through Likert-scale responses from participants, correlating it with test case outcomes (Section~\ref{sec:results-rq1}).
Statistical tests on experimental and survey data are used to investigate the impact of simulators (\beamng vs. \carla), driving views (outside and driver's view), and test case complexity (with/without obstacles/vehicles) on SDC safety perception.

\subsubsection{\RQ{2}: Impact of human interaction on the assessments of SDCs}
Once we know how humans perceive the safety of SDC test cases and how this is related to the OOB metric (\RQ{1}), we investigate whether human-based interactions with the virtual SDC affect the safety perception of the test case.
We argue that the safety perception of a SDC can vary when having the ability to interact, i.e., the possibility to accelerate and deaccelerate the vehicle manually, and previous VR research has shown that interactions can influence the environment positively or negatively~\cite{DBLP:conf/aivr/MolinaJG20,DBLP:conf/chi/PengYLWTYC20,DBLP:conf/vrst/HuangK20,DBLP:conf/vr/HuangLSHK19,DBLP:conf/vr/OgawaNH18,DBLP:conf/vr/HvassLVNNS17,DBLP:conf/vr/RothLGHBLF16,DBLP:journals/ijmms/KwonPC13}. This aspects deserves investigation since it can help developers and researchers in designing better test cases and evaluation metrics, which lead us to our second research question:
\begin{tcolorbox}
\RQ{2}: \rqII
\end{tcolorbox}

\subsubsection{\RQ{3}: Human-based assessment of Realism}
We argue that the level of realism of SDC simulation-based test cases is another important factor influencing the safety perception of SDCs.
It is important to note that the notion of realism relates to the \textit{Reality Gap}~\cite{wang2021exploratory,afzal2021simulation,NgoBR21,RewayHWHKR20} (see Section~\ref{sec:related-work}), which is a critical concern regarding the oracle problem in simulation-based testing: \textit{``due to the different properties of simulated and real contexts, the former may not be a faithful mirroring of the latter"}. 
While recent studies provide solutions for addressing the reality gap, \eg by leveraging domain randomization techniques or using data from real-world observations~\cite{chebotar2019closing,zhang2019adversarial,koos2012transferability}, in the development phase of CPS, there is no prior study that studied and/or characterized the perception of realism of SDC test cases from human participants when 
using VR technologies~\cite{DBLP:conf/aivr/MolinaJG20,DBLP:conf/chi/PengYLWTYC20,DBLP:conf/vrst/HuangK20}.
Hence, to complement \RQ{1} and \RQ{2}, our study addresses 
the following third research question:
\begin{tcolorbox}
\RQ{3}: \rqIII
\end{tcolorbox}
Hence, after the experiments for \RQ{1} and \RQ{2}, we ask the study participants to evaluate the level of realism for \beamng and \carla. Then, we develop a taxonomy of aspects influencing these environments' realism to help improve simulation environments for effective testing of SDCs so that different properties of simulated and real contexts are minimized. 

\subsection{Design overview}
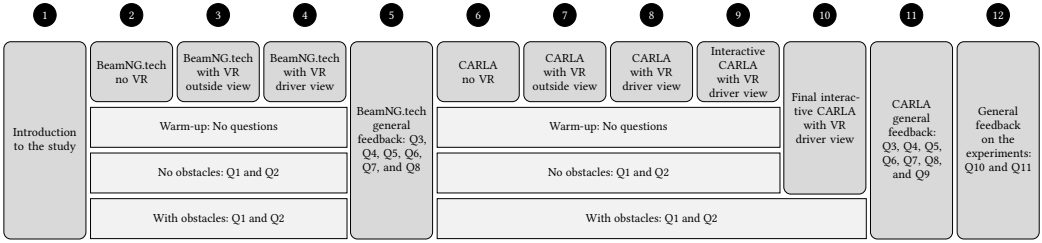
\begin{figure}[t]
    \centering
    \scriptsize
    \resizebox{\linewidth}{!}{\input{imgs/tikz/experiment-survey-single-column}}
    \caption{Design overview with survey question IDs from Table~\ref{tab:survey-questions}}
    \label{fig:experiment-procedure}
\end{figure}

Figure~\ref{fig:experiment-procedure} overviews the design of our study involving 12 steps:

In step \noindent\textbf{\circled{1}}, we welcome and introduce the study participant by explaining the context and the procedure for the experiments. The participant in step \noindent\textbf{\circled{2}} sits before a computer screen and experiences three simulation-based test cases with the \beamng simulator. While sitting before a computer, the participant wears a VR headset for the next steps.
In step 
\noindent\textbf{\circled{3}}, 
the participant experiences three test cases with the \beamng simulator observing the SDC from an \emph{outside view} perspective while in step \noindent\textbf{\circled{4}}, the participant experiences three test cases with the \beamng simulator from a \emph{driver view} perspective.
The step \noindent\textbf{\circled{5}} focuses on general feedback on the experiments with the \beamng simulator.
Then, the steps \noindent\textbf{\circled{2}},\noindent\textbf{\circled{3}},\noindent\textbf{\circled{4}}
are repeated for the \carla simulator in \noindent\textbf{\circled{6}}, \noindent\textbf{\circled{7}}, \noindent\textbf{\circled{8}}.
In step \noindent\textbf{\circled{9}}, for the \carla simulator, the participant, while wearing a VR headset from a driver's view, experiences three test cases in which they can control the SDC speed with a keyboard.
In addition to step \noindent\textbf{\circled{9}}, 
 one group of participants in step \noindent\textbf{\circled{10}}  will experience a crash with the SDC.
 The step \noindent\textbf{\circled{11}} focuses on general feedback on the experiments with the \carla simulator while the step \noindent\textbf{\circled{12}} focuses on general feedback on the overall study.

For the steps \circled{2} - \circled{4}, and \circled{6} - \circled{9}, the participant experiences three test cases.
The first test case is the warm-up so that the participant can familiarize himself or herself with the simulation environment.
The second test case has no obstacles, and the third test case has obstacles (i.e., has higher complexity).
At step \circled{10}, the participant only experiences the complex test case with obstacles.

\subsection{Design implementation}\label{sec:sudy-context}
We implement our design by conducting experiments with our test runner called \framework.
The test runner uses three distinct test cases created by a test generator (see Section~\ref{sec:test-generator-and-runners}).
The participants give responses to our survey questionnaires using \textit{Google Forms}.


\subsubsection{Test cases}
We use three distinct test cases generated by the \textit{Frenetic} test generator \cite{CastellanoCTKZA21} for different purposes.
The first test case is the warm-up that lets the participant familiarize with the simulation environment and view setting, e.g., to get used to the VR headset and the simulator.
Hence, no survey question for this first warm-up test case is provided.
The second test case does not have obstacles, while the third involves obstacles (higher complexity).


\subsubsection{\framework}\label{sec:approach}
We extend the existing test runner \sdcscissor (see Section~\ref{sec:test-generator-and-runners}) 
by implementing \framework (\frameworklong).
Specifically, we implement an interface to run test cases with the \carla simulator for the steps \circled{6} - \circled{10}.
As for \beamng, with \framework we can also add obstacles to the test cases in \carla to achieve similar complexity levels for the experiments.
Additionally,  with \framework, and for steps \circled{9}-\circled{10}, the participants could control the SDC speed with the keyboard.

Test cases generated are processed differently between \beamng and \carla since \carla.  
An automatically generated test case in \beamng (Section \ref{sec:background}) consists of a sequence of XY-coordinates (i.e., the road points).
The \carla simulator, however, does not need all the road points defined in the test.
\framework segments road definitions, using only the start and end points of the segments to declare scenarios in \carla{}. Moreover, it enables user immersion and safety evaluation by automatically adapting test case specifications for \carla and utilizing VR headsets for immersive experiences in its virtual environment.

\subsubsection{Survey questionnaires}
\begin{table}[t]
    \centering
    \newcommand{\ls}[0]{LS\xspace}
    \newcommand{\oa}[0]{OA\xspace}
    \newcommand{\ynm}[0]{SC\xspace}
    \scriptsize
    \caption{Survey questions with Likert-scale (\ls), Open answer (\oa), and Single-choice (\ynm) types}
    \label{tab:survey-questions}
    \begin{tabular}{clc}
    \toprule
    \textbf{ID} & \textbf{Question} & \textbf{Type} \\
    \midrule
    Q1  & What is the perceived safety of the Scenario? & \ls \\
    Q2  & Justify the perceived safety of the Scenario. & \oa \\
    Q3  & How would you scale the realism of scenarios generated by test cases in the simulator? & \ls \\
    Q4  & Justify the level of realism of scenarios generated by test cases. & \oa \\
    Q5  & How would you scale the driving of AI of the simulator? & \ls \\
    Q6  & Justify the driving of AI from the simulator. & \oa \\
    Q7  & How would you scale overall experience with the simulator? & \ls \\
    Q8  & Justify overall experience with the simulator. & \oa \\
    Q9  & How do you compare safety with and without interaction? & \oa \\
    Q10 & Did this experiment change the way you thought about the safety of self-driving cars? & \ynm \\
    Q11 & Please write in a few words on your experience and suggestions. & \oa \\
    \bottomrule
    \end{tabular}
\end{table}

We employ \textit{Google Forms} for our questionnaires, a free and user-friendly survey tool. Table \ref{tab:survey-questions} summarizes participant questions, having multiple choice (MC), open answer (OA), and Likert scale (LS) questions (with values from 1-5, where 1 for very unsafe, 5 for very safe, and 3 for neutral).
to address our research questions (RQs).
Participants answered Q1 and Q2 after the second and third test cases, respectively, with the first test case serving as a warm-up without safety assessment. For Q3-Q8, participants provide responses after all three simulator test executions, i.e., at step \circled{5} for \beamng and step \circled{11} for \carla. Note that at step \circled{11}, we include an additional question, Q9, for experiments involving \carla, which includes interactive scenarios requiring keyboard inputs to control the SDC's speed.



\subsubsection{Experimental Setting}
We conducted experiments in a dedicated, soundproof room to eliminate external distractions. Participants sat at a table equipped with a desktop computer, laptop, and a VR headset. They used the laptop running the \textit{Google Forms} application to complete survey questionnaires and the desktop computer for non-VR experiments. For VR experiments, participants used the \vive headset, known for its high visual resolution, powered by the \textit{nVidia GeForce RTX 3080} and \textit{Windows 10} operating system. Additional extensions were employed to allow a full VR experience to participants, such as \vorpx for \beamng's VR support and the \textit{DReyeVR} extension for \carla, were used. We also integrated \framework to facilitate testing with both \beamng and \carla simulators. 
Furthermore, the participants were allowed to interact with specific SDC test cases, with the keyboard enabling them to adjust the SDC's speed.

\subsection{Study participants}\label{sec:survey}
We recruit participants via email invitations sent to our industrial partners, university students, and researchers across departments. We target various mailing lists, including non-computer science organizations, and leverage social media platforms such as Twitter and LinkedIn. We use physical and digital flyers to attract diverse participants, ensuring a broad range of backgrounds and education levels.

\subsubsection{Pre-survey}\label{sec:pre-survey}
When participants sign up for our experiments, we email them a pre-survey created with \textit{Google Forms} to collect demographic information. This survey includes an introduction to the topic, an overview of the experiment (including approximate time and location), and a recommendation to wear contact lenses. It also provides details about the simulator and VR headset used. Furthermore, the pre-survey includes a disclaimer regarding confidentiality and anonymity and a warning about potential VR-related accidents or fatalities that the participants could experience. Following this section, we gather background information on participants, as detailed in the Appendix (appx.) of our replication package (Section~\ref{sec:data-availability}). These questions cover testing and driving experience, VR technology usage, age, and gender. This additional information helps us investigate potential confounding factors affecting safety and realism perception. 


\subsection{Data collection}
We gather data from two primary sources: the survey (both pre-experiment and during the experiments) and the simulation logs collected during participant experiments.

\subsubsection{Survey data}
For both \beamng and \carla simulators, participants evaluate test cases considering the various questions reported in Table~\ref{tab:survey-questions}.
Specifically, for steps \circled{2} - \circled{4} and \circled{6} - \circled{9}, Likert-scale and text data are collected for each test case except the warm-up case.
For step \circled{10}, only Likert-scale and text data are collected for test cases with obstacles.
Additionally, at steps \circled{5} and \circled{11}, general feedback on the simulators is collected after the test executions with all viewpoints.
Complementary, participants rate the perceived safety and realism of each simulator using Likert-scale values based on their own driving experiences.
Finally, general feedback on the experiments is collected at step \circled{12}. In total, we collected 21 Likert-scale, 23 open, and 1 single-choice response per participant during the experiments.
In addition to the experimental survey, we gather data from the pre-survey (Section~\ref{sec:pre-survey}) to obtain participant demographics, mainly through single-choice and open-text responses.

\subsubsection{Simulation data}
For each test case in each participant's experiment, we collect relevant data, saving logs (see Section~\ref{sec:data-availability}) in JSON files of \framework. These logs include timestamped vehicle position coordinates, sensor data (e.g., fuel, gear, wheel speed), and OOB metric violations (i.e., driving off the lane), categorizing the test as pass or fail based on this metric. Additionally, on \carla, the log structure includes also weather condition details. It is important to note that to enhance our findings further, we also analyze participants' quantitative and qualitative insights both with and without VR headsets as well as when experiencing different driving views.

\subsection{Data analysis}
\subsubsection{\RQ{1} \& \RQ{2}: Perceived level of safety}
We utilize various visualizations, including stacked barplots and boxplots, to assess safety and realism perceptions. We apply statistical tests:
\mannwhitneyu, and \varghadelaney to determine the effective size.
For \RQ{1}, we mainly analyze responses from the test cases where the participant has no interaction with the SDC; for \RQ{2}, we analyze the data where the participant has some direct interactions with the SDC by a keyboard to control the vehicle's speed.
In \RQ{2}, we explore how SDC interactions affect the safety and realism perceptions of participants.
For this, we analyze Likert-scale scores and qualitative feedback.
We employ stacked bar plots to examine data spread across the two categories in steps \circled{8} and \circled{9}.

\subsubsection{\RQ{3}: Taxonomy on realism}

With \RQ{3}, we examine the realism of SDC test cases and their correlation with human safety assessments. We identify and categorize factors affecting test case realism in a taxonomy based on the participant responses in question Q4 at steps \circled{5} and \circled{11}.

We adopt a two-step approach for the initial taxonomy creation. Initially, two authors analyze responses grouped by the simulators: one author focuses on Q4 from step \circled{5} with the \beamng simulator, and the other on Q4 from step \circled{11} with the \carla simulator. Each author proposes categories via an open-card sorting method~\cite{spencer2009card}. In the second step, both authors collaboratively define a meta-taxonomy by discussing their proposed categories. Subsequently, this meta-taxonomy is employed to label all Q4 responses for \beamng and \carla (steps \circled{5} and \circled{11}). To do this, the two authors responsible for the meta-taxonomy and a third author conduct a hybrid card sorting labeling process using online spreadsheets. They individually assign each response to the meta-taxonomy categories or create new categories when necessary. A collaborative approach is employed for validation, where each of the three co-authors reviews and addresses any disagreements in assignments during an online meeting.

%% file: imgs/tikz/experiment-survey-single-column.tex
\tikzstyle{startstop} = [rectangle, text width=15mm, minimum height=12mm, rounded corners, text centered, draw=black, fill=gray!30]
\tikzstyle{process} = [rectangle, minimum width=52mm, minimum height=8mm, text centered, draw=black, fill=gray!10]
\begin{tikzpicture}[node distance=3mm and 5mm]
\node[shape=circle,draw,fill=black,minimum size=13pt, inner sep=0.1pt, outer sep=0pt] (char1) {\textcolor{white}{1}};
\node (Intro) [startstop,below=of char1, text centered, minimum height=39.5mm, anchor=north] {\scriptsize Introduction to the study};
\node[shape=circle,draw,fill=black,minimum size=13pt, inner sep=0.1pt, outer sep=0pt, right=of char1, xshift=8mm] (char2) {\textcolor{white}{2}};
\node (B1) [startstop, below=of char2, xshift=0mm] {\beamng no VR};
\node (B1S0) [process, below=of B1.south west, anchor=north west, yshift=2mm] {Warm-up: No questions};
\node (B1S1) [process, below=of B1S0, yshift=2mm] {No obstacles: Q1 and Q2};
\node (B1S2) [process, below=of B1S1, yshift=2mm] {With obstacles: Q1 and Q2};
\node[shape=circle,draw,fill=black,minimum size=13pt, inner sep=0.1pt, outer sep=0pt, right=of char2, xshift=8mm] (char3) {\textcolor{white}{3}};
\node (B2) [startstop, below=of char3] {\beamng with VR outside view};
%
\node[shape=circle,draw,fill=black,inner sep=0.1pt,minimum size=13pt, outer sep=0pt, right=of char3, xshift=8mm] (char4) {\textcolor{white}{4}};
\node (B3) [startstop, below=of char4] {\beamng with VR driver view};
%
\node[shape=circle,draw,fill=black,inner sep=0.1pt,minimum size=13pt, outer sep=0pt, right=of char4, xshift=8mm] (char5) {\textcolor{white}{5}};
\node (BGeneral) [startstop, below=of char5, text centered, minimum height=39.5mm, xshift=0mm] {\beamng general feedback: Q3, Q4, Q5, Q6, Q7, and Q8};
%
%
%
\node[shape=circle,draw,fill=black,inner sep=0.1pt,minimum size=13pt, outer sep=0pt, right=of char5, xshift=8mm] (char6) {\textcolor{white}{6}};
\node (C1) [startstop, below=of char6, xshift=0cm] {\carla no VR};
\node (C1S0) [process, below=of C1.south west, anchor=north west, yshift=2mm, minimum width=69.5mm] {Warm-up: No questions};
\node (C1S1) [process, below=of C1S0.south west, anchor=north west, minimum width=69.5mm, yshift=2mm] {No obstacles: Q1 and Q2};
\node (C1S2) [process, below=of C1S1.south west, anchor=north west, minimum width=87mm, yshift=2mm] {With obstacles: Q1 and Q2};
\node[shape=circle,draw,fill=black,inner sep=0.1pt, minimum size=13pt, outer sep=0pt, right=of char6, xshift=8mm] (char7) {\textcolor{white}{7}};
\node (C2) [startstop, below=of char7] {\carla with VR outside view};
\node[shape=circle,draw,fill=black,inner sep=0.1pt,minimum size=13pt, outer sep=0pt, right=of char7, xshift=8mm] (char8) {\textcolor{white}{8}};
\node (C3) [startstop, below=of char8] {\carla with VR driver view};
\node[shape=circle,draw,fill=black,inner sep=0.1pt, minimum size=13pt, outer sep=0pt, right=of char8, xshift=8mm] (char9) {\textcolor{white}{9}};
\node (C3I) [startstop, below=of char9] {Interactive \carla with VR driver view};
%
\node[shape=circle,draw,fill=black,inner sep=0.1pt,minimum size=13pt, outer sep=0pt, right=of char9, xshift=8mm] (char10) {\textcolor{white}{10}};
\node (C4I) [startstop, below=of char10, minimum height=30.5mm, text centered] {Final interactive \carla with VR driver view};
\node[shape=circle,draw,fill=black,inner sep=0.1pt, minimum size=13pt, outer sep=0pt, right=of char10, xshift=8mm] (char11) {\textcolor{white}{11}};
\node (CGeneral) [startstop, below=of char11, text centered, minimum height=39.5mm] {\carla general feedback: Q3, Q4, Q5, Q6, Q7, Q8, and Q9};
\node[shape=circle,draw,fill=black,inner sep=0.1pt, minimum size=13pt, outer sep=0pt, right=of char11, xshift=8mm] (char12) {\textcolor{white}{12}};
\node (General) [startstop, below=of char12, minimum height=39.5mm, text centered] {General feedback on the experiments: Q10 and Q11};

\end{tikzpicture}

%% file: sections/03-results.tex
\section{Results}\label{sec:results}
In this section, we present the survey results for \RQ{1}, focusing on participants' safety perception of the test cases, and \RQ{2}, examining how this perception changes when participants can interact with the SDC. For \RQ{3}, we developed a taxonomy by classifying participants' comments on test case realism.

\subsection{\RQ{1}: Human-based assessment of safety  metrics}\label{sec:results-rq1}
To address \RQ{1}, we analyzed Likert scale values across various data subgroups. These subgroups included comparisons between test outcomes (failures and successes based on OOB metrics) and different test case complexities (with and without obstacles). This allowed us to identify factors influencing perceived safety among participants. We present boxplots and statistical tests (appx. B.1) for each subgroup.

\subsubsection{Safety perception of failing vs. passing test cases}

\begin{figure}[t]
    \centering
    \resizebox{!}{4cm}{\input{imgs/tikz/rq1-boxplot-failing-passing-tests}}
    \resizebox{!}{4cm}{\input{imgs/tikz/passing-and-failing-tests-grouped-by-complexity-boxplot}}
    \caption{Perceived safety of failing and passing tests grouped by scenario's complexity}
    \label{fig:rq1-1}
    \label{fig:rq1-2}
\end{figure}
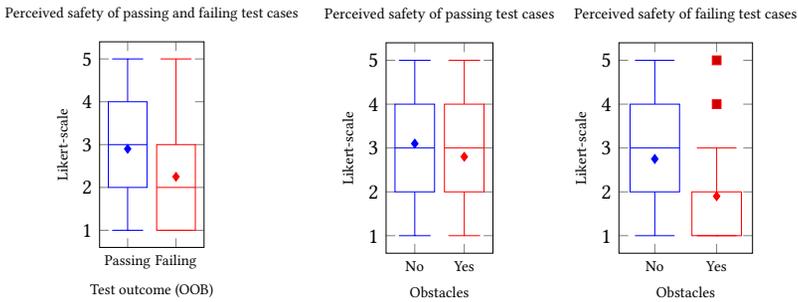

Figure~\ref{fig:rq1-1} illustrates perceived safety distributions for test cases grouped by test outcome (OOB metric). We found a significant difference (Table~\ref{tab:rq1-statistical-test-results}) in how participants rate safety for failing and passing test cases on a Likert scale.

\begin{finding}\label{find:rq1-1}
The passing test cases (i.e., the cases where the OOB metric is not violated) have a higher perception of safety from the participants than those failing (OOB metric is violated).
\end{finding}

The aforementioned Finding~\ref{find:rq1-1} is somewhat expected and is aligned with comments from study participants (appx. C.1). 
These comments pertain to the \beamng simulator, excluding VR and obstacles. We selected these comments for their exclusive focus on SDC lane-keeping, providing qualitative insights into the OOB metric without obstacle influence. Notably, among comments where the SDC violates the OOB metric (test case failure), safety concerns are recurrent:
\textit{``As the car did not drive all the time on the street, I felt unsafe. [...].''- (P3/B1/S1)}''; 
\textit{``When the car starts to go off the road when driving in a curve, it feels pretty unsafe.''- (P31/B1/S1)}''; 
\textit{``Not Very Safe since the car sometimes drove a bit from the road.''- (P45/B1/S1)}.

On passing test cases where the OOB metric is not violated, we can find that the participants gave consistent comments in terms of safety:
\textit{``The car was driving in lane and at a safe speed considering the road is empty.'' - (P16/B1/S1)}; 
\textit{``The car was following the path in a safe way and was not speeding up too much.'' - (P25/B1/S1)}.

All comments that support Finding~\ref{find:rq1-1} are listed in appx. C.1. 

\subsubsection{Safety perception With and Without obstacles}

Additionally, participants assessed test cases with varying complexity, including additional obstacles. Figure~\ref{fig:rq1-2} displays differences in perceived safety, with statistical significance reported in appx. B.1. 
Concretely, failing test cases are generally seen as less safe, but those with added obstacles are perceived as even less safe. 
In contrast to passing test cases, perceived safety remains largely unaffected by the higher complexity of scenarios (e.g., additional obstacles).
As shown in appx. B.1, 
no significant statistical differences were observed in the samples, leading us to conclude:

\begin{finding}\label{find:failing-passing-complexity}\label{find:rq1-2}
There is no statistical difference in safety perception between scenarios with and without obstacles when the OOB metric is not violated. However, when the car goes out of bounds, the scenario is perceived as significantly less safe with obstacles ($p=3.52*10^{-16}$).
\end{finding}

From participants, we received qualitative support for Finding~\ref{find:failing-passing-complexity}. For those feeling unsafe with scene obstacles, here are representative answers:
\textit{``The car crashed toward an obstacle and even running over bumps was not so smooth as humans would do. Definitively more unsafe than the previous scenario.''- (P1/B1/S2)}; 
\textit{``Ran off the road in a curve and hit obstacles without slowing down, which resulted in flat tires.''- (P24/B1/S2)}.

In participants who felt safe or neutral when obstacles were present, consistent comments were reported:
\textit{``It car was running smooth with obstacles, there was a moment when it was too close to one of the obstacle'' - (P16/B1/S2)}; 
\textit{``The vehicle does well to avoid obstacles while maintaining the safe speed'' - (P18/B1/S2)}; 
\textit{``The driver accelerated over all the obstacles and did not have a perfect finish.'' - (P40/B1/S2)}; 
\textit{``Car was driving well. Only at the end it went off the road, but there was no object it bumped into.'' - (P45/B1/S2)}.

All comments that support Finding~\ref{find:failing-passing-complexity} are reported in appx. C.1. 

\subsubsection{Safety perception, with VR and without VR}
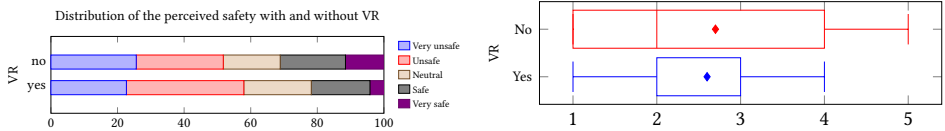
\begin{figure}[t]
    \begin{subfigure}{0.45\linewidth}
        \input{imgs/tikz/rq1-vr-vs-novr-stacked-bar}
    \end{subfigure}
    \begin{subfigure}{0.45\linewidth}
        \input{imgs/tikz/rq1-vr-vs-novr-boxplot}
    \end{subfigure}
        \vspace{-2mm}
    \caption{VR vs. no VR}
    \label{fig:rq1-VR-vs-no-VR-stacked-bar}
        \vspace{-2mm}
\end{figure}

To assess the impact of VR on safety perception, we categorized data into \textit{with VR} and \textit{without VR} groups.
Appx. B.1 
shows no statistically significant difference. However, Figure~\ref{fig:rq1-VR-vs-no-VR-stacked-bar} reveals that \textit{without VR} has more \textit{very unsafe} and \textit{very unsafe} responses. This is also evident from the smaller interquartile range in \textit{with VR} (compared to the \textit{without VR}). 

\begin{finding}\label{find:VR-vs-no-VR}\label{find:rq1-3}
The utilization of VR had a minor impact on safety perception. However, participants using VR tend to perceive scenarios as somewhat less safe, though this difference was not statistically significant (\mannwhitneyu test, $p=0.16$).
\end{finding}

Certain participant comments support Finding~\ref{find:VR-vs-no-VR}. For instance, a \neutral participant stated:
\textit{``The prespective doesnt change much with the vr'' - (P22/B2/S1)}.
Another example is a comment from a participant who felt \veryunsafe:
\textit{``The same as without the VR glasses. The car was not able to keep the middle of the lane and was driving badly compared to a human.'' - (P28/B2/S1)}.

\subsubsection{Different views with different complexity}
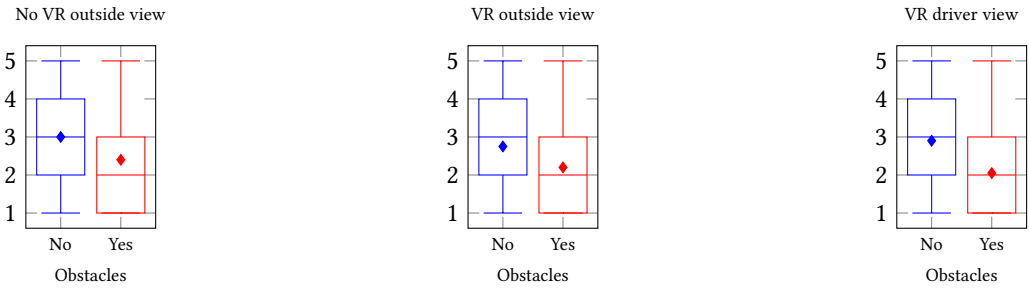
\begin{figure}[t]
    \centering
    \input{imgs/tikz/rq1-different-views-different-complexity-boxplots}
    \vspace{-2mm}
    \caption{Different VR-related views grouped by scenario's complexity}
    \label{fig:rq1-3}
    \vspace{-2mm}
\end{figure}

In Figure~\ref{fig:rq1-3}, we note a decrease in test case safety perception across various viewpoints.
Statistical differences are evident in appx. B.1, 
supporting the following general finding:

\begin{finding}\label{find:different-view-with-different-complexity}\label{find:rq1-4}
Overall, participants found the test cases less safe with obstacles.
\end{finding}

Participants' general comments during the experiment for each simulator qualitatively support Finding~\ref{find:different-view-with-different-complexity}. Representative comments on \beamng driving behavior include:
\textit{``It did not look at safety lines, which is very dangerous if other traffic is involved. It also ran off the road multiple times, which can easily lead to a loss of control. Also, the car rashed into easily avoidable obstacles.'' - (P24/B)}; 
\textit{``At least the AI seems to have an understanding of the general elements of the simulation, like the road. However, it seems to struggle with bumps in the middle of the road and also seems to drive too fast in curvy situations.'' - (P31/B)}.

In the case of \carla, we got the following representative comments on the driving behavior with regard to different complexity of the scenario: 
\textit{``Except at the roundabouts, the car followed traffic rules, signals, and speed limits. However, it kept crashing and losing control in the roundabouts.'' - (P27/C)}; 
\textit{``In most scenarios, the AI did well. From what I have seen during the simulations, it is not able to drive around roundabouts and does not stop at stop signs.'' - (P31/C)}; 
\textit{``very slow driving, unsmooth behavior, always too close to roundabout and abrupt stopping in front  of obstacles.'' - (P41/C)}.


We observe that the perception of safety drops when increasing the complexity (i.e., adding obstacles to the scenario). This observation is coherent among both simulators, \beamng and \carla, as reported by the participants during the experiment.


\subsection{\RQ{2}: Impact of human interaction on the assessments of SDCs}\label{sec:results-rq3}
\begin{figure}
    \centering
    \begin{subfigure}{0.49\linewidth}
        \centering
        \input{imgs/tikz/rq2-interactive-vs-non-interactive-stacked-bar}
    \end{subfigure}
    \begin{subfigure}{0.49\linewidth}
        \centering
        \input{imgs/tikz/rq2-interactive-grouped-by-complexity-stacked-bar}
    \end{subfigure}
    \begin{subfigure}{0.49\linewidth}
        \centering
        \input{imgs/tikz/rq2-non-interactive-grouped-by-complexity-stacked-bar}
    \end{subfigure}
    \vspace{-3mm}
    \caption{Safety perception with and without interaction with the SDC (grouped by complexity)}
    \label{fig:with-and-no-interaction}
\end{figure}
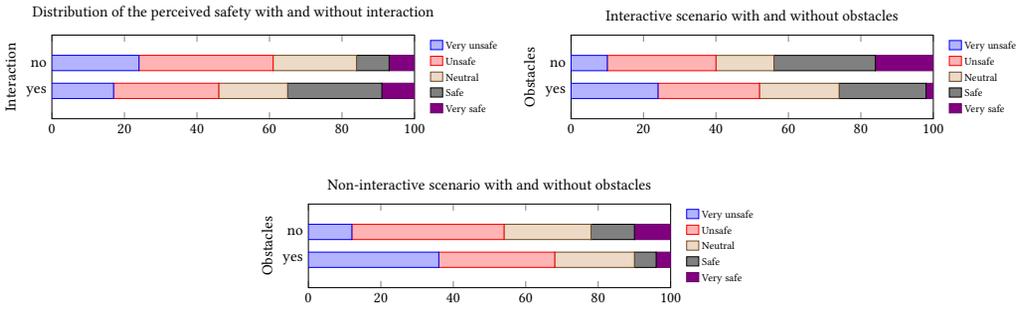

To assess the safety perception of test cases with human interaction with the SDC, participants controlled SDC speed during the test execution. Figure~\ref{fig:with-and-no-interaction} shows the Likert scale of responses. We compare responses when participants can or cannot control the car and when obstacles are present.

\subsubsection{Safety perception with and without interaction with the SDC}
In general, interacting with the SDC enhances participants' perception of safety. From appx. B.2, 
we observe a statistically significant difference, leading to the following finding:
\begin{finding}\label{find:rq2-1}
Safety perception of test cases is not static: When users can interact with the SDC, participants feel significantly safer ($p=0.013$) compared to when they cannot.
\end{finding}

The participants' justification supports Finding\ref{find:rq2-1}, e.g., controlling the SDC speed enhances safety perception, as P1 reported:
\textit{``The fact I could control the car when needed gave me a safer perception of the driving experience. Moreover, I could speed up the car when I wanted to.'' - (P1)}.
However, not all participants perceive interaction-based test cases as inherently safe. For instance, participant P4 comments:
\textit{``With a bit of control, it feels safer, especially being able to adjust the speed in dangerous situations. However, it is still not safe since the car ends up going off-road at the end of the scenario.'' - (P4)}. 
While the SDC remains self-steering, it may still crash despite having speed control capability.

\subsubsection{Safety perception for with and without obstacles}
When interactive test cases involve obstacles, participants perceive them as less safe than obstacle-free scenarios, a statistically significant difference, leading to the following finding:
\begin{finding}\label{find:rq2-2}
Incorporating obstacles into the simulation, where participants interact with the SDC, leads to significantly lower perceived safety in test cases ($p=0.026$) compared to obstacle-free interactive scenarios.
\end{finding}

This finding is also coherent with the answers of the study participants, \eg by P4:
\textit{``It felt safer, especially since it was stopping the speed when it had another car in front.
However, it still went to the footpath, making it not safe'' - (P4)}.
From the comment, we observe safer perception through speed control. P20 also states:
\textit{``it could have stopped before hitting the camion'' - (P20)}.

However, as the study participant cannot control the SDC's steering, some accidents remain unavoidable, as reported by P19:
\textit{``Hit the bike driver'' (P19)}.
P40 gives a clearer comment:
\textit{``Two matters: 1) driver keeps its distance to the can in the front, but with sharp breaks instead of slowing down the car.
2) unable to avoid strange behaviors and drove next to a car with unstable drive and had an accident'' (P40)}.
The participant can maintain distance by adjusting speed, but accidents can occur during lane changes.

In non-interactive test cases, obstacles induce insecurity among participants. However, the level of how they feel unsafe when obstacles are included is higher in the case where the participants can interact with the SDC. This leads to the following finding:
\begin{finding}\label{find:rq2-3}
In the simulation, obstacles in non-interactive SDC test cases reduce safety perception ($p=0.013$). Yet, the ability to interact with the car raises more discomfort (making participants feel less safe) when obstacles are present.
\end{finding}

Besides the statistical tests, we also note participant comments supporting Finding~\ref{find:rq2-3}. Some express discomfort in obstacle scenarios without the ability to control the car, as evident in the following example:
\textit{``The car was breaking and accelerating a lot while being behind the other car, and also the other car was not behaving safely on the road, ending the simulation with an accident between the two, so it felt quite unsafe overall.'' (P25)}.
Some participants also experience the worst-case scenario without control, as reported by P28:
\textit{``It drove extremely close up to the ambulance car and finally crashed into it. therefore, the worst case happens.'' (P28)}.

\subsection{\RQ{3}: Taxonomy on realism}
\newcommand{\catone}{World Objects}
\newcommand{\cattwo}{Dynamics}
\newcommand{\catthree}{Road}
\newcommand{\catfour}{Traffic Elements}
\newcommand{\catfive}{Rule System}
\newcommand{\catsix}{Immersion}
\newcommand{\catseven}{Others}

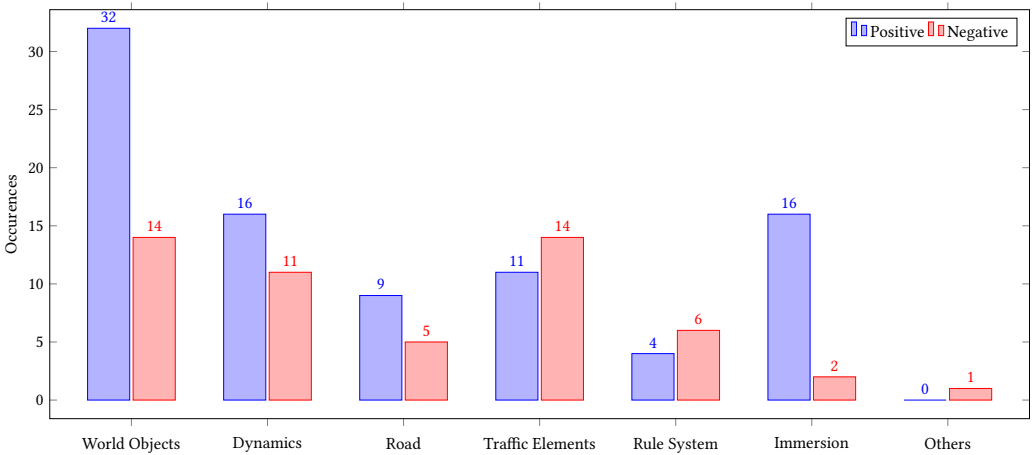
\begin{figure}
    \centering
    \resizebox{\textwidth}{!}{\input{imgs/tikz/taxonomy-realism}}
    \caption{Taxonomy of positive and negative factors impacting the perceived test cases' realism. }
    \label{fig:taxonomy-on-realism}
\end{figure}

\begin{table}[t]
    \centering
    \scriptsize
    \caption{Taxonomy description including  \# of positive and negative comments on the perception of realism}
            \vspace{-1mm}
    \label{tab:taxonomy-description}
    \begin{tabular}{lp{7cm}ccc}
        \toprule
        \multirow{2}{*}{\textbf{Category}} & \multirow{2}{*}{\textbf{Description}} & \multicolumn{3}{c}{\textbf{Occurrences}} \\ \cline{3-5} 
        & & \textbf{Positive} & \textbf{Negative} & \textbf{Total} \\ \midrule
        \catone & This category relates to comments of participants on the accuracy of visual looks and design of all elements in the virtual environment, such as the weather, landscape, car design, traffic objects, etc., and how the graphical resolution is perceived. & 32 & 14 & 46 \\
        \cattwo & This category relates to participants' comments on the physical dynamics of the elements in the virtual environment. For example, if the movement of the cars is physically realistic and reasonable or if crashes are realistically simulated from a physical perspective. & 16 & 11 & 27 \\
        \catthree & This category relates to participants' comments on the road itself; to what extent the shape, surface, and structure are reasonably expected in the real world. & 9 & 5 & 14 \\
        \catfour & This category relates to participants' comments on the placement of the elements in the virtual environment. Furthermore, this category considers comments on the location and scale of the placed elements but also the quantity of the elements. & 11 & 14 & 25 \\
        \catfive & This category relates to participants' comments on the traffic laws and the common sense of humans for resolving certain issues in specific traffic situations. A car should, for example, stop at a red signal and stop signs. Furthermore, the car should not drive recklessly and avoid dangerous situations (e.g., driving too close to other vehicles). & 4 & 6 & 10 \\
        \catsix & This category relates to participants' comments on the immersive experiences. It applies to comments where participants express their feelings on how they experience the virtual environment and how they acoustically, visually, physically, and haptically sense it. & 16 & 2 & 18 \\
        \catseven & This category relates to participants' comments that do not fit into the above categories. & 0 & 1 & 1 \\
        \bottomrule
        \vspace{-3mm}
    \end{tabular}
\end{table}


Realism is a crucial aspect to consider when evaluating test case \textit{safety}. We created a taxonomy to gauge the perceived realism of study participants. Two coders used open card sorting on 50 comments each to establish categories, which were later reviewed by a third coder. Table~\ref{tab:taxonomy-description} presents the seven resulting categories with their descriptions.

Next, two coders independently classified 100 comments using the designed taxonomy. Disagreements were resolved by a third coder. Table~\ref{tab:taxonomy-description} and Figure~\ref{fig:taxonomy-on-realism} show the classification of comments related to question Q4 in steps \circled{5} and \circled{11}. We categorized comments as \textit{positives} (increasing realism) and \textit{negatives} (decreasing realism) in the taxonomy.
We observe that most classifications fall under \textit{World Objects}, totaling 46, with 32 positives and 14 negatives.
\begin{finding}\label{find:rq3-1}
Several factors (e.g., the surroundings, car design, and object scale) impact the participants' perceived realism.
The \textit{World Objects} category dominates 
with 32 \textit{positive} (e.g., car design)  and 14 \textit{negative} (e.g., traffic objects) aspects affecting realism perception. 
\end{finding}

Examples of positive comments with the \beamng simulator:
\textit{``The realism is quite good, especially in the car design.
The car structure was damaged after crashing; the wheels were getting broken, and there was smoke coming out.
The inside view of the car was also pretty real, with the driver's hand moving the steering wheel and all the car panel commands.
[...]
.'' - (B/P4)}; 
\textit{``They respect the scale from the objects.'' - (B/P22)}.
Examples of positive comments for the \carla simulator:
\textit{``The surroundings have more detail, which made it feel more realistic.'' - (C/P31)}; 
\textit{``The environment (lighting, obstacles) feels quite real.'' - (C/P17)}. 
An example of a negative comment:
\textit{``The grass, the horizon as well, and the red vertical lines do not look very realistic.'' - (B/P3)}.
Besides finding in Section~\ref{find:rq3-1}, we noted that the \textit{Immersion} category generally received positive comments about perceived realism.
\begin{finding}\label{find:rq3-2}
The \textit{Immersion} category primarily comprises comments on factors that affect realism (e.g., view, perspective). It includes 16 \textit{positive} (e.g., the realism of driver's seat) and 2 \textit{negative} (e.g., low realism outside the vehicle) comments influencing participants' perceived realism.
\end{finding}
This finding is reasonable since a driver sits in the driver's seat, unlike the perspective in a video game. The following quotes support this:
\textit{``The driver seat simulator felt very realistic.'' - (B/P14)}; 
\textit{``It was different when I sat in the car than from outside, so it felt more real. But still looked like a game, so not that realistic.'' - (B/P21)}. 
In summary, comments on \textit{Immersion} were positive, indicating that the driver seat viewpoint and VR usage enhanced perceived realism.

%% file: imgs/tikz/rq1-boxplot-failing-passing-tests.tex
\begin{tikzpicture}
\begin{axis}[
xticklabels={\scriptsize Passing, \scriptsize Failing},
y=0.8cm,
boxplot/draw direction=y,
x=0.8cm,
y=0.7cm,
ylabel={\scriptsize Likert-scale},
ytick={1,2,3,4,5},
xtick={1,2},
title={\scriptsize Perceived safety of passing and failing test cases},
xlabel={\scriptsize Test outcome (OOB)}
]
\addplot+ [
    boxplot prepared={
    lower whisker=1, lower quartile=2,
    median=3,
    average=2.9,
    upper quartile=4, upper whisker=5,
    },
] coordinates {};
\addplot+[
    boxplot prepared={
    lower whisker=1, lower quartile=1,
    median=2,
    average=2.25,
    upper quartile=3, upper whisker=5,
    },
] coordinates {};
\end{axis}
\end{tikzpicture}

%% file: imgs/tikz/passing-and-failing-tests-grouped-by-complexity-boxplot.tex
\begin{tikzpicture}
\begin{axis}[boxplot/draw direction=y, x=0.8cm, y=0.7cm, ylabel={\scriptsize Likert-scale}, ytick={1,2,3,4,5}, xtick={1,2}, xlabel={\scriptsize Obstacles}, xticklabels={\scriptsize No, \scriptsize Yes}, xticklabel style = {align=center, rotate=0}, title={\scriptsize Perceived safety of passing test cases}]
    \addplot+ [boxplot prepared={lower whisker=1, lower quartile=2, median=3, upper quartile=4, upper whisker=5, average=3.1}] coordinates {};
    \addplot+ [boxplot prepared={lower whisker=1, lower quartile=2, median=3, upper quartile=4, upper whisker=5, average=2.8}] coordinates {};
\end{axis}
\end{tikzpicture}
\hfill
\begin{tikzpicture}
\begin{axis}[boxplot/draw direction=y, x=1cm, y=0.7cm, ylabel={\scriptsize Likert-scale}, ytick={1,2,3,4,5}, xtick={1,2}, xlabel={\scriptsize Obstacles}, xticklabels={\scriptsize No, \scriptsize Yes}, xticklabel style = {align=center, rotate=0}, title={\scriptsize Perceived safety of failing test cases}]
    \addplot+ [boxplot prepared={lower whisker=1, lower quartile=2, median=3, upper quartile=4, upper whisker=5, average=2.75}] coordinates {};
    \addplot+ [boxplot prepared={lower whisker=1, lower quartile=1, upper quartile=2, upper whisker=3, average=1.9}] coordinates { (2,4) (2,5) };
\end{axis}
\end{tikzpicture}

%% file: imgs/tikz/rq1-vr-vs-novr-stacked-bar.tex
\resizebox{\linewidth}{!}{
\begin{tikzpicture}
\begin{axis}[
title=Distribution of the perceived safety with and without VR,
xbar stacked,
xmin=0, xmax=100,
ylabel={VR},
ylabel style={draw=white},
symbolic y coords={yes,no},
height=3.5cm,
ytick=0pt,
width=10cm,
enlarge y limits=1,
ytick=data,
legend pos=outer north east,
legend cell align=left,
legend columns=1,
legend style={draw=none},
]
\addplot coordinates {(22.67,yes) (25.63,no)};
\addplot coordinates {(35.26,yes) (26.13,no)};
\addplot coordinates {(20.25,yes) (17.09,no)};
\addplot coordinates {(17.63,yes) (19.6,no)};
\addplot coordinates {(5.28,yes) (11.56,no)};
\legend{\scriptsize Very unsafe, \scriptsize Unsafe, \scriptsize Neutral, \scriptsize Safe,\scriptsize Very safe}
\end{axis}
\end{tikzpicture}
}

%% file: imgs/tikz/rq1-vr-vs-novr-boxplot.tex
\resizebox{\linewidth}{!}{
\begin{tikzpicture}
\begin{axis}[
boxplot/draw direction=x,
ytick={1,2},
yticklabels={\scriptsize Yes, \scriptsize No},
ylabel={\scriptsize VR},
y=0.8cm,
]
\addplot+ [
    boxplot prepared={
    lower whisker=1, lower quartile=2,
    average=2.6,
    upper quartile=3, upper whisker=4,
    },
] coordinates { };
\addplot+[
    boxplot prepared={
    lower whisker=1, lower quartile=1,
    median=2,
    average=2.7,
    upper quartile=4, upper whisker=5,
    },
] coordinates {};
\end{axis}
\end{tikzpicture}
}

%% file: imgs/tikz/rq1-different-views-different-complexity-boxplots.tex
\begin{tikzpicture}
\begin{axis}[
boxplot/draw direction=y,
title={\scriptsize No VR outside view},
xtick={1,2},
ytick={1,2,3,4,5},
xticklabels={\scriptsize No, \scriptsize Yes},
xlabel={\scriptsize Obstacles},
x=0.8cm,
y=0.5cm,
]
\addplot+ [
    boxplot prepared={
    lower whisker=1, lower quartile=2,
    median=3,
    average=3,
    upper quartile=4, upper whisker=5,
    },
] coordinates { };
\addplot+[
    boxplot prepared={
    lower whisker=1, lower quartile=1,
    median=2,
    average=2.4,
    upper quartile=3, upper whisker=5,
    },
] coordinates {};
\end{axis}
\end{tikzpicture}
\hfill
\begin{tikzpicture}
\begin{axis}[
boxplot/draw direction=y,
title={\scriptsize VR outside view},
xtick={1,2},
ytick={1,2,3,4,5},
xticklabels={\scriptsize No, \scriptsize Yes},
xlabel={\scriptsize Obstacles},
x=0.8cm,
y=0.5cm,
]
\addplot+ [
    boxplot prepared={
    lower whisker=1, lower quartile=2,
    median=3,
    average=2.75,
    upper quartile=4, upper whisker=5,
    },
] coordinates { };
\addplot+[
    boxplot prepared={
    lower whisker=1, lower quartile=1,
    median=2,
    average=2.2,
    upper quartile=3, upper whisker=5,
    },
] coordinates {};
\end{axis}
\end{tikzpicture}
\hfill
\begin{tikzpicture}
\begin{axis}[
boxplot/draw direction=y,
title={\scriptsize VR driver view},
xtick={1,2},
ytick={1,2,3,4,5},
xticklabels={\scriptsize No, \scriptsize Yes},
xlabel={\scriptsize Obstacles},
x=0.8cm,
y=0.5cm,
]
\addplot+ [
    boxplot prepared={
    lower whisker=1, lower quartile=2,
    median=3,
    average=2.9,
    upper quartile=4, upper whisker=5,
    },
] coordinates { };
\addplot+[
    boxplot prepared={
    lower whisker=1, lower quartile=1,
    median=2,
    average=2.05,
    upper quartile=3, upper whisker=5,
    },
] coordinates {};
\end{axis}
\end{tikzpicture}

%% file: imgs/tikz/rq2-interactive-vs-non-interactive-stacked-bar.tex
\resizebox{\linewidth}{!}{
\begin{tikzpicture}
\begin{axis}[
title=Distribution of the perceived safety with and without interaction,
xbar stacked,
xmin=0, xmax=100,
ylabel={Interaction},
ylabel style={draw=white},
symbolic y coords={yes,no},
height=3.5cm,
ytick=0pt,
width=10cm,
enlarge y limits=1,
ytick=data,
legend pos=outer north east,
legend cell align=left,
legend columns=1,
legend style={draw=none},
]
\addplot coordinates {(17,yes) (24,no)};
\addplot coordinates {(29,yes) (37,no)};
\addplot coordinates {(19,yes) (23,no)};
\addplot coordinates {(26,yes) (9,no)};
\addplot coordinates {(9,yes) (7,no)};
\legend{\scriptsize Very unsafe, \scriptsize Unsafe, \scriptsize Neutral, \scriptsize Safe,\scriptsize Very safe}
\end{axis}
\end{tikzpicture}
}

%% file: imgs/tikz/rq2-interactive-grouped-by-complexity-stacked-bar.tex
\resizebox{\linewidth}{!}{
\begin{tikzpicture}
\begin{axis}[
title=Interactive scenario with and without obstacles,
xbar stacked,
xmin=0, xmax=100,
ylabel={Obstacles},
ylabel style={draw=white},
symbolic y coords={yes,no},
height=3.5cm,
ytick=0pt,
width=10cm,
enlarge y limits=1,
ytick=data,
legend pos=outer north east,
legend cell align=left,
legend columns=1,
legend style={draw=none},
]
\addplot coordinates {(24,yes) (10,no)};
\addplot coordinates {(28,yes) (30,no)};
\addplot coordinates {(22,yes) (16,no)};
\addplot coordinates {(24,yes) (28,no)};
\addplot coordinates {(2,yes) (16,no)};
\legend{\scriptsize Very unsafe, \scriptsize Unsafe, \scriptsize Neutral, \scriptsize Safe,\scriptsize Very safe}
\end{axis}
\end{tikzpicture}
}

%% file: imgs/tikz/rq2-non-interactive-grouped-by-complexity-stacked-bar.tex
\resizebox{\linewidth}{!}{
\begin{tikzpicture}
\begin{axis}[
title=Non-interactive scenario with and without obstacles,
xbar stacked,
xmin=0, xmax=100,
ylabel={Obstacles},
ylabel style={draw=white},
symbolic y coords={yes,no},
height=3.5cm,
ytick=0pt,
width=10cm,
enlarge y limits=1,
ytick=data,
legend pos=outer north east,
legend cell align=left,
legend columns=1,
legend style={draw=none},
]
\addplot coordinates {(36,yes) (12,no)};
\addplot coordinates {(32,yes) (42,no)};
\addplot coordinates {(22,yes) (24,no)};
\addplot coordinates {(6,yes) (12,no)};
\addplot coordinates {(4,yes) (10,no)};
\legend{\scriptsize Very unsafe, \scriptsize Unsafe, \scriptsize Neutral, \scriptsize Safe,\scriptsize Very safe}
\end{axis}
\end{tikzpicture}
}

%% file: imgs/tikz/taxonomy-realism.tex
\resizebox{\linewidth}{!}{
\begin{tikzpicture}
\begin{axis}[
ybar,
bar width=25pt,
width=22cm,
height=10cm,
nodes near coords,
enlargelimits=0.05,
enlarge x limits=0.1,
legend style={at={(0.9,0.98)},
anchor=north,legend columns=-1},
ylabel={Occurences},
symbolic x coords={\catone,\cattwo,\catthree,\catfour,\catfive,\catsix,\catseven},
xtick=data,
x tick label style={yshift=-0.1cm,rotate=0,anchor=north},
]
\addplot+[ybar] plot coordinates { (\catone,32) (\cattwo,16) (\catthree,9) (\catfour,11) (\catfive,4) (\catsix,16) (\catseven,0) };
\addplot+[ybar] plot coordinates { (\catone,14) (\cattwo,11) (\catthree,5) (\catfour,14) (\catfive,6) (\catsix,2)  (\catseven,1) };
\legend{Positive,Negative}
\end{axis}
\end{tikzpicture}
}



%% file: sections/04-discussion.tex
\section{Discussion}\label{sec:discussion}
We 
first discuss safety considerations for simulation-based tests, including \RQ{1} and interactive test cases \RQ{2}. Then, we delve into realism by discussing the taxonomy of influencing factors.

\subsection{\RQ{1} \& \RQ{2}: Human-based safety assessment of simulation-based test cases}
The study participants perceived passing test cases (OOB metric not violated) as safer than failing ones (Finding~\ref{find:rq1-1}), aligning with the OOB metric-based test oracle. This observation is supported by~\cite{DBLP:conf/icst/Jahangirova0T21}, where participants' assessment of driving quality correlates with metrics related to the SDC's lateral position. The OOB metric generally reflects test case safety. However, the extent to which the safety perception varies depending on certain simulation factors (e.g., obstacle inclusion) remains unclear. Hence, we conducted experiments with test cases featuring additional obstacles.

In Section~\ref{find:rq1-2}, we found that adding obstacles to a passing test case does not significantly affect safety perception. However, participants perceive failing test cases as less safe with additional obstacles. 
Therefore, human safety perception does not proportionally align with the OOB metric. The OOB metric can be violated, but it still does not distinguish the case if there are additional obstacles in the test case, but the human does and perceives the test case unsafer.

We experimented with different immersion levels (i.e., various viewpoints), and as reported in Finding~\ref{find:rq1-3}, participants using VR headsets perceived test cases as slightly less safe. This perception change is minimal when evaluating VR. Consequently, when using humans as oracles, outcomes vary based on immersion levels in virtual environments. Hence, similar human-based studies on simulation-based test cases for SDCs \cite{DBLP:conf/icst/Jahangirova0T21} may exhibit a slight bias if immersion is not considered. When grouping safety perceptions of test cases by their assessed viewpoints, cases with obstacles were generally perceived as less safe than those without obstacles (Finding~\ref{find:rq1-4}). Thus, using the OOB metric as an oracle may not always accurately represent safety perceptions from a human perspective. This observation aligns with the example 
illustrated by Figure~\ref{fig:01-motivating-example} and Figure~\ref{fig:02-motivating-example}.

As shown in Finding~\ref{find:rq2-1}, participants perceived test cases as safer when they could control the vehicle's speed (i.e., they express a higher trust level in the SDC behavior), which means that the safety perception of simulation-based test cases depends on the user interaction levels. Having control over the vehicle impacts safety perception, which may not align with the OOB metric. In the case of test cases involving participant interaction, safety perception generally decreases when obstacles are present, as indicated by Finding~\ref{find:rq2-2}. This aligns with the findings for non-interactive test cases, as highlighted in Finding~\ref{find:rq2-3}.

\subsection{\RQ{3}: Taxonomy on test cases' realism}

As shown in Finding~\ref{find:rq3-1}, most participants' comments on Question Q4 fall under the \textit{World Objects} category.
As discussed in Section~\ref{sec:introduction}, we conjecture that assessing test case safety should also consider realism.
The importance of \textit{World Objects}, with respect to realism, confirms the fact that pure lane-keeping (as it is the focus of OOB) is not enough for doing a realistic safety assessment.
Given that most comments related to test case realism are categorized as \textit{World Objects}, it becomes essential to prioritize when evaluating test case safety.
The \textit{Immersion} category predominantly features comments expressing a positive or heightened sense of realism, as revealed in Finding~\ref{find:rq3-2}.
Participants' immersion, particularly their viewpoint, influences perceived realism.
Notably, the driver seat perspective yields a higher realism perception, as evident in comments on Finding~\ref{find:rq3-2}, consequently impacting safety perception.
The importance of immersion, with respect to realism, confirms that static 2D assessment (again, as it is the focus for OOB) is not enough for doing a realistic safety assessment.

When we take a closer look at the participants' demographics and how they assess the level of realism, we observed that the participants in the age range between 18 and 30 years tend to assess the test cases 17\% more realistically (Likert scale) than the older participants.
Another insight is that we do not observe a different assessment of realism among the genders.
Hence, there are confounding factors that influence the perception of realism, such as the age of the participant.
This aspect suggests that the reality-gap characteristics are not deterministic measures as they depend on the human perception that might vary, as for the case of the participants'age.

\subsection{Implications \& Lessons learned}\label{sec:lessons-learnt}
The oracle definition for SDCs is many-fold as the safety has different aspects characterizing it.
The OOB metric may not always reflect human safety perception in test cases due to various unaccounted factors.
To enhance simulation-based testing, SDC testers and practitioners should consider devising alternative metrics that better align with human safety perception.
Interacting with the car boosts perceived safety, potentially due to distrust in the AI driving the SDC.
Future research should explore this further, ruling out other influencing factors.
If low trust in AI is the main issue, this suggests shaping the direction of autonomous driving research toward increasing the level of trustworthiness of SDCs, which represents an important limiting factor to SDC real-world adoption.

As motivated in Section~\ref{sec:introduction}, realism significantly influences the safety perception of SDCs, as reflected in participants' comments on Q4.
For this reason, we have created a taxonomy of factors that affect realism in simulation-based SDC testing, to guide future research in the field.
The taxonomy provides an overview of factors impacting the realism of SDC simulation-based testing.
We argue that our taxonomy is instrumental in supporting future research on the \textit{ perceived reality-gap}, which is critical to bridge the gap between the simulation-based outcome of a test case and what happens eventually in the real world.
Furthermore, we think the taxonomy provides a base for investigating similar limitations in other CPS application domains, which leverage simulation environments and target to improve the human perception of the realism and safety of CPSs. 
%
%

%% file: sections/05-threats-to-validity.tex
\section{Threats to validity}\label{sec:threats-to-validity}

\subsection{Threats to internal validity}
%
The study participants rated safety and realism based on their immersion into the scenario. To limit the risks of unbiased assessments, we employed modern VR technology (\vive) to enhance immersion. The simulators, \beamng and \carla, utilize distinct predefined maps. \beamng employs a flat map from the SBST tool competition~\cite{DBLP:conf/sbst/PanichellaGZR21}, while \carla uses built-in urban-like maps, which impose some constraints on road definition. These differing maps may lead to varying perceptions of test case safety and realism due to their distinct natures. This is something we plan to investigate for future work. 

The different personal interactions with the study participants might influence the participants' focus during the experiments. To limit this risk, we used a protocol sheet during the experiments to ensure that all steps of the experiments were equally performed to minimize this threat.

\subsection{Threats to external validity}
We recruited study participants primarily from an academic computer science background, which may not represent the general population.
To address this potential bias, we ensured diversity in terms of age, gender, and driving experience, reducing the influence of factors beyond professional background.
Another concern is the focus on the OOB metric, which may introduce bias as there are various metrics for evaluating SDCs in simulation environments.
We chose OOB due to its widespread use among researchers and practitioners, as documented in recent studies~\cite{BeamNGresearch,DBLP:journals/tosem/BirchlerKDPP23, DBLP:conf/icst/Jahangirova0T21,DBLP:conf/sbst/PanichellaGZR21,GambiJRZ22}.
Our study's limited use of only two simulators, \beamng and \carla, restricts the generalizability of our findings to these specific platforms.
However, we selected them because they are widely adopted in academia and industry, ensuring the reproducibility of our results compared to less-maintained options such as Udacity\footnote{https://github.com/udacity/self-driving-car} and SVL~\cite{rong2020lgsvl}.

%% file: sections/06-related-work.tex
\section{Related work}\label{sec:related-work}
In this section, we elaborate on related work on testing in virtual environments and assessing the quality of oracles in the context of CPS.
We group the recent and ongoing research concerning topics that are relevant to our investigation such as
\begin{inparaenum}[(i)]
    \item simulation-based testing,  
    \item the testing metrics adopted, the oracle problem,  and   
    \item VR in software engineering. 
\end{inparaenum}

\subsection{Simulation-based testing}
The automated testing of Cyber-Physical Systems (CPSs) remains an ongoing research challenge~\cite{DiSorboTOSEM2023}.
In this context, simulation-based testing emerges as a promising approach to enhance testing practices for Safety-Critical Systems (SDC)~\cite{PiazzoniCAYSV21,NguyenHG21,DBLP:journals/tosem/BirchlerKDPP23,DBLP:journals/ese/BirchlerKBGP23} and to support test automation~\cite{AlconTAC21, Wotawa21,afzal2021simulation, afzal2018crashing, wang2021exploratory}. 
Past research on testing CPS in simulation environments focused on monitoring CPS and predicting unsafe states \cite{DiSorboTOSEM2023,10.1145/3377811.3380353} of the systems using simulation environments~\cite{DBLP:conf/icst/XuAY21,10.1145/3377811.3380353} as well as generating scenarios programmatically~\cite{DBLP:conf/icse/ParkJBC20} or based on real-world observations~\cite{DBLP:conf/sigsoft/GambiHF19,StoccoPT23}. 
Recent research also proposed cost-effective regression testing techniques, including test selection~\cite{DBLP:journals/ese/BirchlerKBGP23}, prioritization~\cite{DBLP:journals/tosem/BirchlerKDPP23,DBLP:journals/jss/ArrietaWSE19} and minimization techniques to expose CPS faults or bugs earlier in the development and testing process.
This research effort fundamentally contributed toward more robust and reliable simulation-based testing practices.
However, it remains challenging to replicate the same bugs observed in physical tests within simulations~\cite{wang2021exploratory, afzal2020study} and generate representative simulated test cases that uncover realistic bugs~\cite{afzal2021simulation}.
Hence, previous research in the field was conducted on the premise that simulation environments sufficiently represent, with high fidelity, safety-critical aspects of the real world according to human judgments.
In our paper, we hypothesize that the current simulation-based testing of SDCs (and general CPSs) does not always align with the human perception of safety and realism, which heavily impacts the effectiveness of simulation-based testing in general.
To that end, in our research, we investigated when and why the safety metrics of simulation-based test cases of SDCs match human perception.


\subsection{Testing metrics \& the Oracle Problem}
To automatically infer the expected test outcome from a given input remains an unsolved challenge, which is known as the oracle problem.
Many research papers propose some techniques to address this problem into the context of traditional software systems such as generating oracles~\cite{DBLP:conf/wcre/ArrietaOHSAA22} or improving already existing test oracles~\cite{DBLP:conf/issta/JahangirovaCHT16,DBLP:conf/sigsoft/TerragniJTP20,DBLP:conf/gecco/TerragniJPT21,DBLP:conf/icse/TerragniJTP21}.
In either case, the previous research do not show an approach that produce fully optimal and effective oracles.
However, while the oracle problem still remains an open challenge which requires humans to define the oracle, for the sake of test automation, several code coverage and mutation score metrics have been proposed for for quantitatively assessing the quality of traditional software systems.

Software engineering for CPS is increasingly explored, with recent efforts mainly focused on bug characterization~\cite{DBLP:conf/icse/GarciaF0AXC20}, testing~\cite{DBLP:journals/tecs/DeshmukhHJMP17, DBLP:conf/icse/AbdessalemNBS18, DBLP:conf/icse/ZhouLKGZ0Z020}, and verification~\cite{DBLP:conf/icse/ChowdhurySJC20} of self-adaptive CPSs.  
Another emerging area of research is related to the automated generation of oracles for testing and localizing faults in CPSs based on simulation technologies.
For instance, Menghi {\em et al.}~\cite{DBLP:conf/sigsoft/MenghiNGB19} proposed SOCRaTes, an approach to automatically generate online test oracles in Simulink able to handle CPS Simulink models featuring continuous behaviors and involving uncertainties.
The oracles are generated from requirements specified in a signal logic-based language. 
In this context, for the sake of test automation, just like traditional software testing, simulation-based testing of SDCs relies on an oracle that determines whether the observed behavior of a system under test is safe or unsafe. 
To that end, current research on automated safety assessment focuses primarily on a limited set of temporal and non-temporal safety metrics for SDCs \cite{10.1145/3579642,DBLP:conf/sbst/PanichellaGZR21,GambiJRZ22,DBLP:journals/ese/BirchlerKBGP23}.  
In particular, the out-of-bound (OOB) non-temporal metric is largely adopted for assessing SDCs in simulation-based testing~\cite{NguyenHG21,GambiJRZ22,DBLP:conf/sbst/PanichellaGZR21}, to determine if a test case fails or passes.
However, it is yet unclear whether this metric serves as a meaningful oracle for assessing the safety behavior of SDCs in simulation-based testing in general.

This study is built on our hypothesis that current simulation-based testing of SDCs does not always align with the human perception of safety and realism, and for this reason, we focus on understanding and characterizing this mismatch in our research.
Close to our work, a recent study~\cite{DBLP:conf/icst/Jahangirova0T21} conducted a human-based study and observed that correlations between the computed quality metrics and the perceived quality by humans are meaningful for assessing the test quality for SDCs.
However, such previous work did not investigate the factors that define the test quality and realism of the simulation environments from a human point of view with the use of virtual reality~\cite{DBLP:conf/hri/SilveraBA22} as done in our work.

A critical concern concerning the oracle problem in simulation-based testing is represented by the \emph{Reality Gap}~\cite{wang2021exploratory,afzal2021simulation,NgoBR21,RewayHWHKR20}.
Due to the different properties of simulated and real contexts, the former may not be a faithful mirroring of the latter.
Simulations are necessarily simplified for computational feasibility yet reflect real-world phenomena at a given level of veracity, the extent of which is the result of a trade-off between accuracy and computational time~\cite{collins2020traversing}.
Robotics simulations rely on the replication of phenomena that are difficult to accurately replicate, e.g., simulating actuators (i.e., torque characteristics, gear backlash), sensors (i.e., noise, latency), and rendered images (i.e., reflections, refraction, textures).
This gap between reality and simulation is commonly referred to as the \textit{reality-gap}\cite{collins2020traversing}.
A closely related problem concerns the concrete realistic \emph{bug reproduction} and exposure in simulation environments~\cite{wang2021exploratory,afzal2021simulation}.
It is indeed challenging to capture the same bugs as physical tests~\cite{wang2021exploratory,afzal2020study} and to \emph{generate effective test cases} that can expose real-world bugs in simulation~\cite{afzal2021simulation}.
While recent studies provide solutions for addressing the reality gap (e.g., leveraging domain randomization techniques or using data from real-world observations)  \cite{chebotar2019closing,lee2020camera,zhang2019adversarial,koos2012transferability,salvato2021crossing,collins2020traversing} in the development phase of CPS, there is no prior study that investigated and/or characterized the perception of realism of SDC test cases from human participants.
This study focuses on addressing this specific open question in the context of RQ3.
\subsection{Immersion Technology in Software engineering} 
Furthermore, using VR for software engineering was also considered by~\cite{DBLP:conf/sigsoft/HoffN021, DBLP:conf/kbse/MehraSKPB20} but with another focus as well.
They used VR to gain design knowledge from legacy systems by using diferent visualization approaches using immersion technologies.
Furthermore, most papers~\cite{DBLP:conf/icse/MehraSKPB20,DBLP:conf/icse/SharmaMKP19,DBLP:conf/icse/SharmaMKP18} referring to the potential use of VR and AR for the workspace of software development teams.
In general, the use of VR and AR in software engineering is not well studied yet, and the only papers available or mainly vision papers for future research~\cite{DBLP:conf/wcre/MerinoL020}.
However, in our work, we present a practical application of VR for assessing the test oracles with a Human-in-the-Loop approach.



%% file: sections/07-conclusion.tex
\section{Conclusion}\label{sec:conclusion}
Testing self-driving car (SDC) software, such as traditional software, relies on safety and quality oracles.
However, depending solely on metrics such as the OOB for simulation-based SDC testing can be limited in terms of reliability and perceived realism from a human standpoint.
In this study, we explored when and why safety metrics align with human perception in SDC testing.
We conducted an empirical study with 50 participants from diverse backgrounds, evaluating their perception of test case safety and realism.
We observed that the safety perception of SDC significantly decreases as test case complexity rises.
Interestingly, safety perception improves when participants can control the SDC's speed, indicating that OOB metric is not sufficient to match/model human (more subjective) factors.
Additionally, realism perception varies with the complexity of scenarios (i.e., object additions) and different participant viewpoints.
These findings emphasize the need for more meaningful safety metrics that align with human perception of \textit{safety} and \textit{realism} to bridge the current problem of the \textit{reality-gap} in simulation-based testing.
Future work should also consider other safety metrics, as suggested by recent studies~\cite{10.1145/3579642}, to enhance SDC software testing in simulation environments and improve safety and realism.

\section{Data Availability}\label{sec:data-availability}
A replication package with data, code, and appendices is publicly available on Zenodo~\cite{zenodo-replication-package}.

%% file: sections/09-appendices.tex
\appendix
\section{Study participants}
\begin{table}[h!]
\caption{Education level of participants}
\label{table:summarizes_participants}
\scriptsize
\centering
\begin{tabular}{l>{\centering}p{1cm}>{\centering}p{1cm}>{\centering}p{1cm}>{\centering}p{1cm}>{\centering}p{1cm}>{\centering}p{1cm}>{\centering\arraybackslash}p{1cm}}
\toprule
\multirow{2}{*}{\textbf{Field/Profession}}  & \multicolumn{6}{c}{\textbf{Education level}}                                                                                                                                                                      & \multirow{2}{*}{\textbf{Total}}  \\
                                              \cline{2-7} \vspace{-2mm} \\
                                            & \textit{Senior}       & \multirow{2}{*}{\textit{Postdoc}}     & \multirow{2}{*}{\textit{PhD}}     & \multirow{2}{*}{\textit{Master}}  & \multirow{2}{*}{\textit{Bachelor}}    & \textit{other}                    &   \\
                                            & \textit{researcher}   &                                       &                                   &                                   &                                       & \textit{education}           & \\ \midrule    
Artificial intelligence (AI)                & -                     & -                                     & -                                 & 1                                 & -                                     & -                                 & 1\\
AI ethics / Political science               & -                     & -                                     & 1                                 & -                                 & -                                     & -                                 & 1 \\    
Biology                                     & -                     & 1                                     & 2                                 & -                                 & -                                     & -                                 & 3 \\ 
Business administration                     & -                     & -                                     & -                                 & -                                 & 2                                     & -                                 & 2\\   
Computer science                            & 1                     & 2                                     & 5                                 & 22                                & 6                                     & 1                                 & 37 \\ 
Robotics                                    & -                     & -                                     & -                                 & 1                                 & -                                     & -                                 & 1 \\    
Mechanical engineering                      & -                     & -                                     & -                                 & -                                 & 1                                     & -                                 & 1 \\    
Industrial engineering                      & -                     & -                                     & -                                 & 1                                 & -                                     & -                                 & 1 \\    
Architecture                                & -                     & -                                     & -                                 & 1                                 & -                                     & -                                 & 1 \\    
Law                                         & -                     & -                                     & 1                                 & -                                 & -                                     & -                                 & 1 \\    
Commercial clerk                            & -                     & -                                     & -                                 & -                                 & -                                     & 1                                 & 1 \\    
\midrule
\textbf{Total}                              & 1                     & 3                                     & 9                                 & 26                                & 9                                     & 2                                 & \textbf{\participantscount} \\
\bottomrule
\end{tabular}
\end{table}

\begin{table*}[h!]
    \centering
    \scriptsize
    \caption{(with \textit{High Educ.} refers to Faculty and/or Senior Researcher)}
    \resizebox{0.8\linewidth}{!}{
    \begin{tabular}{llrrcl}
    \toprule
    \textbf{Participant} & \textbf{Gender} & \textbf{Age} & \textbf{Driving Experience} & \textbf{Field/Profession} & \textbf{Educational Level}  \\ \midrule
    P1  & man   & 31-35 & \textgreater10 years & Computer Science                                           & Higher Professional Education                                \\
    P2  & man   & 26-30 & \textgreater10 years & Computer Science                                           & Masters                                                      \\
    P3  & man   & 31-35 & \textgreater10 years & Computer Science                                           & Masters                                                      \\
    P4  & man   & 26-30 & 6-10 years & Computer Science                                           & PhD                                                         \\
    P5  & woman & 26-30 & 6-10 years & Computer Science                                           & Masters                                                     \\
    P6  & woman & 26-30 & 1-2 years & Computer Science                                           & Masters                                                      \\
    P7  & man   & 31-35 & 3-6 years & Computer Science                                           & Masters                                                      \\
    P8  & man   & 26-30 & 1-2 years & Computer Science                                           & Masters                                                      \\
    P9  & man   & \textgreater55   & \textgreater10 years & Computer Science                                           & Bachelors                                                    \\
    P10 & woman & 31-35 & 3-6 years & Computer Science                                           & Masters                                                      \\
    P11 & man   & 31-35 & \textgreater10 years & Computer Science                                           & Bachelors                                                    \\
    P12 & man   & 26-30 & 3-6 years & Business administration/ Banking and Finance/ Economics    & Bachelors                                                    \\
    P13 & man   & 18-25 & 3-6 years & Computer Science                                           & Bachelors                         \\
    P14 & woman & 18-25 & less than one year & Computer Science                                           & Masters                                             \\
    P15 & man   & 18-25 & 3-6 years & Computer Science                                           & Masters                                                      \\
    P16 & man   & 18-25 & 6-10 years & Computer Science                                           & Masters                                                     \\
    P17 & man   & 26-30 & 3-6 years & Computer Science                                           & Masters                                                      \\
    P18 & man   & 26-30 & 3-6 years & Computer Science                                           & Masters                                                      \\
    P19 & man   & 26-30 & 6-10 years & Computer Science                                           & Masters                                                     \\
    P20 & man   & 26-30 & 6-10 years & Computer Science                                           & PhD                                                         \\
    P21 & woman & 18-25 & 1-2 years & Business administration/ Banking and Finance/ Economics    & Bachelors                                                    \\
    P22 & woman & 18-25 & 3-6 years & Computer Science                                           & Bachelors                                                    \\
    P23 & man   & 26-30 & 3-6 years & Computer Science                                           & Masters                                                      \\
    P24 & man   & 18-25 & 3-6 years & Computer Science                                           & Masters                                                      \\
    P25 & woman & 26-30 & less than one year & Biology                                           & Postdoc                  \\                                   
    P26 & man     & 18-25   & 3-6 years & Computer Science & PhD \\
    P27 & man     & 31-35              & \textgreater10 years                                          & Computer Science & Masters \\
    P28 & man     & 31-35              & \textgreater10 years                                          & Computer Science & Masters \\
    P29 & man     & 18-25             & 3-6 years                                          & Computer Science & Masters \\
    P30 & man     & 31-35             & \textgreater10 years                                          & Robotics & Masters \\
    P31 & man     & 26-30              & \textgreater10 years                                          & Computer Science & Masters \\
    P32 & man     & 26-30              & 3-6 years                                          & Computer Science & Bachelors \\
    P33 & woman   & 26-30              & 3-6 years                                          & Computer Science & Postdoc \\
    P34 & man     & 26-30              & 3-6 years                                          & Computer Science & Masters \\
    P35 & man     & 18-25              & 3-6 years                                          & Computer Science & Masters \\
    P36 & man     & 26-30              & 3-6 years                                          & Computer Science & PhD \\
    P37 & woman   & 26-30              & 6-10 years                                          & AI ethics / Political science & PhD \\
    P38 & man     & \textgreater55                & \textgreater10 years                                          & Computer Science & Postdoc \\
    P39 & man     & 26-30     & 1-2 years                                          & Computer Science & Masters \\
    P40 & man     & 31-35              & \textgreater10 years                                          & Computer Science & PhD \\
    P41 & woman   & 31-35             & less than one year                                          & Artificial Intelligence & Masters \\
    P42 & man     & 18-25              & 3-6 years                                          & Mechanical Engineering & Bachelors \\
    P43 & woman   & 26-30              & 6-10 years                                          & Computer Science & other \\
    P44 & woman   & 31-35             & 1-2 years                                          & Industrial engineering & Masters \\
    P45 & woman   & 31-35             & less than one year                                          & Architecture & Masters \\
    P46 & woman   & 31-35              & \textgreater10 years                                          & Law & PhD \\
    P47 & woman   & 51-55              & \textgreater10 years                                          & Commercial Clerk & other \\
    P48 & woman   & 26-30              & 6-10 years                                          & Biology & PhD \\
    P49 & woman   & 18-25              & 6-10 years                                          & Biology & PhD \\
    P50 & woman   & 26-30     & less than one year                                          & Computer Science & Bachelors \\
\bottomrule

    \end{tabular}}
    \label{tab:participant-overview}
\end{table*}

\clearpage
\section{Statistical results on safety perception}
\subsection{\RQ{1}}\label{sec:stat-results-rq1}
\begin{table}[h!]
\centering
\caption{Statistical test results for \RQ{1}}
\resizebox{!}{8cm}{
\begin{tabular}{lp{2cm}<{\centering}r}
    \toprule
    \multirow{2}{*}{\textbf{Method}} & \multirow{2}{*}{\textbf{Test statistic}} & \textbf{p-value} \\
     & & \textbf{Effect size $A_{12}$} \\ \midrule
     \multicolumn{3}{l}{\underline{\emph{Failing vs. Passing (Figure~\ref{fig:rq1-1}):}}} \\
     \multirow{2}{*}{\shapirowilk} & $0.86$ & ** $3.69*10^{-18}$ \\
                                  & $0.90$  & ** $4.41*10^{-13}$ \\
    \varghadelaney &  & $0.34$ \\
    \mannwhitneyu & $40461.5$ & ** $6.2*10^{-14}$ \vspace{5mm} \\
     \multicolumn{3}{l}{\underline{\emph{Failing: No obstacles vs. with obstacles (Figure~\ref{fig:rq1-2}):}}} \\
     \multirow{2}{*}{\shapirowilk} & $0.90$ & ** $4.71*10^{-9}$ \\
                                  & $0.79$  & ** $2.15*10^{-16}$ \\ 
    \varghadelaney &  & $0.71$ \\
    \mannwhitneyu & $26507.0$ & ** $3.52*10^{-16}$ \vspace{5mm} \\ 
    \multicolumn{3}{l}{\underline{\emph{Passing: No obstacles vs. with obstacles (Figure~\ref{fig:rq1-2}):}}}  \\
     \multirow{2}{*}{\shapirowilk} & $0.89$ & ** $5.08*10^{-10}$ \\
                                  & $0.90$  & ** $1.56*10^{-7}$ \\ 
    \varghadelaney &  & $0.56$ \\
    \mannwhitneyu & $12719.5$ & $0.06$ \vspace{5mm} \\ 
     \multicolumn{3}{l}{\underline{\emph{VR vs. No VR (without interactive scenarios) (Figure~\ref{fig:rq1-VR-vs-no-VR-stacked-bar}):}}}  \\
     \multirow{2}{*}{\shapirowilk} & $0.89$ & ** $1.77*10^{-16}$ \\
                                  & $0.88$  & **$2.26*10^{-11}$ \\ 
    \varghadelaney &  & $0.47$ \\
    \mannwhitneyu & $36817.5$ & $0.16$ \vspace{5mm} \\ 
     \multicolumn{3}{l}{\underline{\emph{No VR outside view: No obstacles vs. with obstacles (Figure~\ref{fig:rq1-3} and~\ref{fig:rq1-VR-vs-no-VR-stacked-bar}):}}}  \\
     \multirow{2}{*}{\shapirowilk} & $0.89$ & ** $5.48*10^{-7}$ \\
                                  & $0.82$  & **$1.02*10^{-9}$ \\ 
    \varghadelaney &  & $0.66$ \\
    \mannwhitneyu & $6591.5$ & ** $6.74*10^{-5}$ \vspace{5mm} \\ 
     \multicolumn{3}{l}{\underline{\emph{VR outside view: No obstacles vs. with obstacles (Figure~\ref{fig:rq1-3}):}}}  \\
     \multirow{2}{*}{\shapirowilk} & $0.89$ & ** $4.97*10^{-7}$ \\
                                  & $0.85$  & ** $1.17*10^{-8}$ \\ 
    \varghadelaney &  & $0.65$ \\
    \mannwhitneyu & $6460.0$ & ** $2.075*10^{-4}$ \vspace{5mm} \\ 
     \multicolumn{3}{l}{\underline{\emph{VR driver view: No obstacles vs. with obstacles (Figure~\ref{fig:rq1-3}):}}}  \\
     \multirow{2}{*}{\shapirowilk} & $0.90$ & ** $2.21*10^{-6}$ \\
                                  & $0.83$  & ** $2.23*10^{-9}$ \\ 
    \varghadelaney &  & $0.71$ \\
    \mannwhitneyu & $7082.5$ & ** $1.45*10^{-7}$ \\ 
\bottomrule
\end{tabular}}
\label{tab:rq1-statistical-test-results}
\end{table}

\clearpage
\subsection{\RQ{2}}\label{sec:stat-results-rq2}
\begin{table}[h!]
\centering
\caption{Statistical test results for \RQ{2} on the safety perception with interactive scenarios (*:$\alpha < 5\%$, **:$\alpha < 1\%$)}
\begin{tabular}{lp{2.6cm}<{\centering}r}
    \toprule
    \multirow{2}{*}{\textbf{Method}} & \multirow{2}{*}{\textbf{Test statistic}} & \textbf{p-value} \\
     & & \textbf{Effect size $A_{12}$} \\ \midrule
     \multicolumn{3}{l}{\underline{\emph{Interactive vs. non-interactive scenario:}}} \\
     \multirow{2}{*}{\shapirowilk} & 0.90 & **$1.43*10^{-6}$ \\
                                  & 0.87 & **$9.89*10^{-8}$ \\
    \varghadelaney &  & 0.60 \\
    \mannwhitneyu & 5983.5 & *0.013 \vspace{5mm} \\
     \multicolumn{3}{l}{\underline{\emph{Interactive scenario: No obstacles vs. with obstacles (Figure~\ref{fig:rq1-2}):}}} \\
     \multirow{2}{*}{\shapirowilk} & 0.89 & **0.0003 \\
                                  & 0.88  & **0.0001 \\ 
    \varghadelaney &  & 0.626 \\
    \mannwhitneyu & 1565.0 & *0.026 \vspace{5mm} \\ 
     \multicolumn{3}{l}{\underline{\emph{Non-interactive scenario: No obstacles vs. with obstacles (Figure~\ref{fig:rq1-2}):}}} \\
     \multirow{2}{*}{\shapirowilk} & 0.877 & **$9.022*10^{-5}$ \\
                                  & 0.846 & **$1.223*10^{-5}$ \\ 
    \varghadelaney &  & 0.6392 \\
    \mannwhitneyu & 1598.0 & *0.013 \\ 
    \bottomrule
\end{tabular}
\label{tab:rq2-statistical-test-results}
\end{table}

\clearpage
\section{Comments on safety perception}
\subsection{\RQ{1}: Without interaction}\label{sec:comments-on-safety-perception}
\begin{table*}[h!]
    \centering
    \scriptsize
    \caption{Color-coded comments by safety perception on \beamng (no VR, without obstacles) of participants supporting Finding~\ref{find:rq1-1}}
    \resizebox{0.8\linewidth}{!}{
    \begin{tabular}{llp{10cm}}
    \toprule
    \textbf{Code} & \textbf{Perceived Safety} & \textbf{Comment} \\ \midrule
    P3/B1/S1 & \veryunsafe & ``As the car did not drive all the time on the street I felt unsafe. Especially if there would be some obstacles.'' \\
    P4/B1/S1 & \unsafe & ``While the car managed to detect the road correctly, it deviated a bit from it on a couple occasions. Even though it rightly reached the end, in a real-life scenario deviating from the road could cause a fatality.'' \\
    P5/B1/S1 & \unsafe & ``The car was fast and did not stay on track.'' \\
    P7/B1/S1 & \unsafe & ``Cannot term it safe as car drove off the road and into the grass. But since there was no threat by driving on the grass, I won't term it very unsafe.'' \\
    P11/B1/S1 & \safe & ``The car was on track all the time.'' \\
    P13/B1/S1 & \neutral & ``The car wasn't on the middle of the lane and there were a few moments in which it almost went of but didnt.'' \\
    P16/B1/S1 & \safe & ``The car was driving in lane and at a safe speed considering the road is empty.'' \\
    P17/B1/S1 & \unsafe & ``The car is not lane keeping properly and even seems to steer off the road at the end.'' \\
    P19/B1/S1 & \unsafe & ``Not exactly following the road. Unsafe for narrow passages of roads.'' \\
    P20/B1/S1 & \unsafe & ``It drives on the grass and does the curves with too much speed.'' \\
    P24/B1/S1 & \veryunsafe & ``Ran off the roads multiple times and did not follow safety lines in curves.'' \\
    P25/B1/S1 & \verysafe & ``The car was following the path in a safety way and was not speeding up too much'' \\
    P26/B1/S1 & \unsafe & ``The car went out of the way in at least 2 times.'' \\
    P26/B1/S1 & \veryunsafe & ``The car veered off the path multiple times crossing the hard white and yellow lines. It also seemed too fast while turning.'' \\
    P29/B1/S1 & \unsafe & ``Outside of the lines.'' \\
    P30/B1/S1 & \unsafe & ``Drive out of the road and not in the middle of the lane'' \\
    P31/B1/S1 & \unsafe & ``When the car starts to go off the road when driving in a curve it feels pretty unsafe.'' \\
    P33/B1/S1 & \unsafe & ``it went out of the road twice.'' \\
    P34/B1/S1 & \unsafe & ``exited lane twice.'' \\
    P35/B1/S1 & \unsafe & ``The car cut two corners and was off the road with 2 wheels at a time.'' \\
    P36/B1/S1 & \unsafe & ``car went out of the road partially on curves, could be harmful if the road sides have different level'' \\
    P37/B1/S1 & \safe & ``Overall, I do not think I would be in danger in this scenario, but I would still feel anxious because the car drove overboard and  the wheels were in the grass. If a driver would do that I would ask him or her whether he or she is feeling ok.'' \\
    P40/B1/S1 & \unsafe & ``The car crossed the lines on the both side of the road a few times.'' \\
    P41/B1/S1 & \unsafe & ``slightly got off the road, somewhat crash to pile in the end.'' \\
    P43/B1/S1 & \unsafe & ``Road side was not kept well, so it would seem like the driver would not have full control of the car, had it been a human driving.'' \\
    P44/B1/S1 & \unsafe & ``unsafe - crossing the line and invading the other lane.'' \\
    P45/B1/S1 & \unsafe & ``Not Very Safe since the car sometimes drove a bit from the road.'' \\
    P49/B1/S1 & \neutral & ``Does not drive exactly between the lines, so not super safe.'' \\
    P50/B1/S1 & \unsafe & ``The car went off the lanes multiple times.'' \\
    \bottomrule
    \end{tabular}}
\end{table*}

\clearpage
\begin{table*}[h!]
    \centering
    \scriptsize
    \caption{Color-coded comments by safety perception on \beamng (no VR, with obstacles) of participants supporting Finding~\ref{find:failing-passing-complexity}}
    \resizebox{0.8\linewidth}{!}{
    \begin{tabular}{llp{10cm}}
    \toprule
    \textbf{Code} & \textbf{Perceived Safety} & \textbf{Comment} \\ \midrule
    P1/B1/S2    & \veryunsafe   & ``The car crashed toward an obstacle and even running over bumps was not so smooth as humans would do. Definitively more unsafe that the previous scenario.'' \\
    P3/B1/S2    & \veryunsafe   & ``so the same feeling as in scenario 1 as the car was not able to follow the street proberly. In the end car even crashed into a bin'' \\
    P4/B1/S2    & \unsafe       & ``From an AI standpoint, the car's driving technique was similar to scenario 1. While it managed to follow the road, it deviated from it in some occasions, finally crashing with some obstacles. It also surprised me that the car didn't stop accelerating after crashing. As such, I perceive it as unsafe (I would not use such a car in real life).'' \\
    P7/B1/S2    & \unsafe       & ``Hitting the barricade is what makes the scenario unsafe.'' \\
    P8/B1/S2    & \veryunsafe   & ``The initial drive was good, but the car quickly rammed into the obstacle, potentially coming to a halt and unable to recover, and the accident itself was very rough'' \\
    P9/B1/S2    & \neutral      & ``Eindruck das eher schnell gefahren wurde, am Schluss die Strasse verlassen :-('' \\
    P10/B1/S2   & \veryunsafe   & ``easily distracted by obstacles in the path'' \\
    P13/B1/S2   & \veryunsafe   & ``the car was wobbly. didnt stick to the lane. ran into obstacles and didnt take proper measures to get out of the same'' \\
    P14/B1/S2   & \neutral      & ``Safe because of the dividers. Unsafe for unqualified drivers'' \\
    P15/B1/S2   & \unsafe       & ``Car crashed and the tront wheels and suspension were damaged, also it was going too fast for the bumps'' \\
    P16/B1/S2   & \neutral  & ``It car was running smooth with obstacles, there was a moment when it was too close to one of the obstacle.'' \\
    P17/B1/S2   & \veryunsafe  & ``The car is hitting an obstacle in the road and seems to be unable to progress any further making it very dangerous.'' \\
    P18/B1/S2   & \verysafe &  ``The vehicle does well to avoid obstacles while maintaining the safe speed'' \\
    P19/B1/S2   & \unsafe  & ``It seemed the car was not in control. With the turns and very narrowly escaping the obstacles. Wouldn''t feel safe in it'' \\
    P20/B1/S2   & \veryunsafe  & ``it bumped on an easily avoidable side object and couldn't even get out of it'' \\
    P21/B1/S2   & \veryunsafe  & ``the car hit an obstical and wanted to go on without stopping'' \\
    P24/B1/S2   & \veryunsafe  & ``Ran off the road in a curve and hit obstacles without slowing down, which resulted in flat tires.'' \\
    P25/B1/S2   & \neutral  & ``Car was a bit bumpy along the way and actually hit one of the obstacle on the left even if was controlling the speed well along the way it still felt a bit unsafe'' \\
    P27/B1/S2   & \veryunsafe  & ``The car started too fast and did not seem to have any control over its speed or direction. It crashed onto the pylons multiple times as well.'' \\
    P28/B1/S2   & \veryunsafe  & ``The car did not take the speed bumps into account and was not driving in the middle of the lane. At the very end it even drove completely off...'' \\
    P30/B1/S2   & \veryunsafe  & ``Hit the obstacles in the middle of the lane. Car had an accident.'' \\
    P31/B1/S2   & \veryunsafe  & ``Despite the relatively narrow road, the car was going pretty fast. Therefore, it felt a bit unsafe. The AI even hit a bump which is the main reason that it feels unsafe.'' \\
    P32/B1/S2   & \unsafe  & ``Going fast on bumps and offroad multiple times. Parked offraod at the end.'' \\
    P33/B1/S2   & \veryunsafe  & ``The car went out of the road once and creashed due to one obstracle. It could not recover fom the crash.'' \\
    P34/B1/S2   & \veryunsafe  & ``hit obstacle, fast over bumps'' \\
    P35/B1/S2   & \veryunsafe  & ``Went over the speed bumps to fast and crashed.'' \\
    P36/B1/S2   & \veryunsafe  & ``no speed change on the speed bumps, partiall loss of controll after some hard crush in the end.'' \\
    P37/B1/S2   & \neutral  & ``Still not feeling in danger, but the car went faster and drifted a little bit in the curve whereas driving should be smoother than that.'' \\
    P38/B1/S2   & \unsafe  & ``Touched obstacles'' \\
    P39/B1/S2   & \unsafe  & ``car was trying to avoid obstacle which caused it to go wide on road'' \\
    P40/B1/S2   & \neutral  & ``The driver accelerated over all the obsticles and did not have a perfect finish.'' \\
    P41/B1/S2   & \veryunsafe  & ``collision with obstacles, jumps on road and changes of trajectory'' \\
    P43/B1/S2   & \unsafe  & ``Car was not staying on the road, as well as driving to fast for the given conditions. It was again not stayingon the right side of the road during corners.'' \\
    P44/B1/S2   & \veryunsafe  & ``crashing the obstacle and having a crash'' \\
    P45/B1/S2   & \safe  & ``Car was driving well. Only at the end it went off the road, but there was no object it bumped into.'' \\
    P46/B1/S2   & \veryunsafe  & ``the car was driving too fast and thus hit obstacles. and in order to get back on the road, it went on the wrong side of the road, which made me feel unsafe as a passenger.'' \\
    P48/B1/S2   & \unsafe  & ``car did not hold the line'' \\
    P49/B1/S2   & \veryunsafe  & ``does not drive exactly between the lines and then hit the pole. gives a scary feeling.'' \\
    P50/B1/S2   & \veryunsafe  & ``The car went off the lane and crashed into multiple obstacles. It got stuck at one obstacle and tried to drive through it, which didn't succeed.'' \\
    \bottomrule
    \end{tabular}}
\end{table*}

\clearpage
\subsection{\RQ{2}: With interaction}
\begin{table}[h!]
    \centering
    \scriptsize
    \caption{Interactive scenario without obstacles}
    \resizebox{0.8\linewidth}{!}{
    \begin{tabular}{llp{10cm}}
    \toprule
    \textbf{Code} & \textbf{Perceived Safety} & \textbf{[C3] Justify the perceived safety of the interactive scenario 1?} \\ \midrule
 P1  & Neutral & the fact i could control the car when needed, gave me a safer perception of the driving experience. Moreover, i could speed up the car when i wanted to. \\
 P2  & Safe & I could still stop or slow down the car at the stop signs. \\
 P3  & Very Unsafe & anticipation of bad happening in the same circle from previous scenarios \\
 P4  & Unsafe & With a bit of control it feels safer, especially being able to adjust the speed in dangerous situations. However, it is still not safe since the car ends up going off road at the end of the scenario. \\
 P5  & Safe & it was okay \\
 P6  & Very Unsafe & had accident \\
 P7  & Unsafe & It was unsafe as it ran on the sidewalk of roundabout. \\
 P8  & Safe & Safe overall, nothing to complain about. Even after making it extremely safe, the car followed all the road rules. \\
 P9  & Unsafe & konnte beim Stop die Geschwindigkeit verringern, im Roundabout dann wieder unsicher da zu schnell \\
 P10 & Very Unsafe & car crashed  \\
 P11 & Neutral & the car complted the track without issue \\
 P12 & Neutral & Turing have a big issue. \\
 P13 & Very Unsafe & ran into an accident. the interaction wasnt very responsive hence it made me feel even more unsafe than i usually would \\
 P14 & Safe & Decent control of speed \\
 P15 & Safe & Follows the road well. \\
 P16 & Very Safe & no accident and smooth   \\
 P17 & Unsafe & The car ignore stop markings on the road. \\
 P18 & Neutral & The car follows the traffic rules and speed limit well but misses the lane while turning at the end. \\
 P19 & Unsafe & majoity of the drive was safe. but again hit the sideways while taking left in the roundabout.  \\
 P20 & Safe & it is safer when controlling safety zones \\
 P21 & Safe & it slowed down at the stop sign when i clicked unsafe \\
 P22 & Neutral & It feels better because I feel I have control but still its not as resposibe as I want it to be \\
 P23 & Unsafe & The car does not stop at stop markings. It had problems at the roundabout. \\
 P24 & Unsafe & drove well, but still hit the curb in the roundabout \\
 P25 & Safe & Till the roundabout the scenario was very safe but when it had to turn there the safety and stabilty of the car decreased quite a lot, but overall it was still quite positive experience. \\
 P26 & Neutral & The safety is betteer since I can control the speed of the car. But still the car crashed.  \\
 P27 & Unsafe & It was safe for the most part, except at the end when the car mounted the curb at the round-about. \\
 P28 & Safe & I knew where the pain points are, so I could counteract the bad behaviour of the AI. Nevertheless it crashed.  \\
 P29 & Unsafe & Car didn't respect the stops on its own \\
 P30 & Very Safe & Having control over the speed makes it much safer. \\
 P31 & Very Safe & I was feeling save during the whole time.  \\
 P32 & Unsafe & Car went over the roundabout. \\
 P33 & Unsafe & The car changed lane in middle of the road in the start. It went above the roundabout at the end. \\
 P34 & Neutral & Better than without, but in the end its still close with the curb \\
 P35 & Very Safe & I was able to slow down the car in front of the intersection and in front of the roundabout \\
 P36 & Safe & I could fix the mentioned issues at the first turn with the use of speed limit controll.  \\
 P37 & Safe & Everything was fine. At the end the car still did go on the sidewalk, so not so good, but there was no sound, so it felt not as bad ahah.  \\
 P38 & Very Safe & Influence of the speed is good \\
 P39 & Very Safe & It was safe and easy to control \\
 P40 & Unsafe & I could not stop the agent to move up the square at the end even with my feedback and I felt unsafe for this reason. \\
 P41 & Unsafe & Felt okay in the beginning (still rather too slow), felt unsafe in direction of roundabout, went way too close \\
 P42 & Very Safe & Responsive yet issues with the roundabout. \\
 P43 & Unsafe & Ignored stop sign, crashed into roundabout because driving with too much speed \\
 P44 & Neutral & slowing down before the STOP sign and moderate the speed in the roundabout  \\
 P45 & Unsafe & Did not break at the stop sign and drove too hard at the roundabout where there is no view what comes around the corner. \\
 P46 & Safe & i was able to adjust the speed and make the car stop at stop signs which made me feel safe and gave me back some control.  \\
 P47 & Very Unsafe & - \\
 P48 & Safe & More control feels safer \\
 P49 & Very Safe & felt safe, especially since i can regulate the speed myself. gives control over the car back which is good. especially for roundabouts where everyone has a different preferred speed to go over it. \\
 P50 & Safe & The car drove nicely and steadily. I only reduced the speed in the end when entering the roundabout. \\
    \bottomrule
    \end{tabular}}
\end{table}

\clearpage

\begin{table}[h!]
    \centering
    \scriptsize
    \caption{Interactive scenario with obstacles}
    \resizebox{0.8\linewidth}{!}{
    \begin{tabular}{llp{10cm}}
    \toprule
    \textbf{Code} & \textbf{Perceived Safety} & \textbf{[C3] Justify the perceived safety of the interactive scenario 1?} \\ \midrule
 P1  & Unsafe &  even if i was able to control the car, the vehicle was not able to have smooth driving actions close to the other car (e.g., fast acceleration and quick stop). It is also true that this happen in real time, depending on the driver. The final incidents made the whole scenario unsafe. \\
 P2  & Neutral &  I could not react to all obstacles but the AI could manage it.\\
 P3  & Very Unsafe &  anticipation of bad happening in the same circle from previous scenarios\\
 P4  & Unsafe &  It felt safer, especially since it was stopping the speed when it had another car in front. However, it still went to the foot-path making it not safe.\\
 P5  & Very Unsafe &  the car in the front which was always stopping was annoying\\
 P6  & Unsafe &  had accident\\
 P7  & Very Unsafe &  It collided with the vehicle in the end. SDC should be able to identify these kind of day to day aberrations.\\
 P8  & Safe &  The speed control with ctrl and right enter was good, I was able to control the car not climbing on top of the roundabout at the end.\\
 P9  & Neutral &  Ich habe versucht zu lernen, wie die AI auf meine Befehle betreffs Sicherheitsgefuehl reagiert. Konnte dadurch im Roundabout auch genuegend verzoegern, wodurch im mich sicherer fuehlte.\\
 P10 & Unsafe &  not able to follow path\\
 P11 & Very Unsafe &  the car meet with an accident\\
 P12 & Unsafe &  unstable at turning\\
 P13 & Safe &  was good. felt like i followed the lane properly and the rules too\\
 P14 & Safe &  Decent drive with good speed control\\
 P15 & Safe &  It does foolow the saf/unsafe instructions well.\\
 P16 & Unsafe &  it hit a car \\
 P17 & Unsafe &  The car ignores stop markings on the road, turns too aggressively and veeres of the road.\\
 P18 & Neutral &  The car does a good job of avoiding other vehicles and following traffic lights but a little shaky at the end\\
 P19 & Very Unsafe &  Hit the bike driver\\
 P20 & Unsafe &  it could have stopped before hitting the camion\\
 P21 & Very Unsafe &  at the end I felt verzy unsafe beacause I had the feeling I am not controlling it like in the beginning I coud not do anything\\
 P22 & Neutral &  Also I feel more in control with the Interactive one but I was not able to prevent the crash\\
 P23 & Very Unsafe &  The car was simply not operaing well, I felt very unsafe.\\
 P24 & Safe &  slowed down at stop sign, kept proper distance and I was able to tell the car to slow down\\
 P25 & Unsafe &  The behaviour of the car in the presence of the other car was quite unsafe, with a lot of breaking and accelerating. The behaviour in the absence of the other car but with heavy rain was much better even still having problems dealing with the roundabout.\\
 P26 & Safe &  Having the controll is better for safe perception\\
 P27 & Unsafe &  It was safe for the most part, except at the end when the car mounted the curb at the round-about and lost control.\\
 P28 & Neutral &  I was finally able to stop at the stop sign and but still not able to avoid the crash. Therefore overall safe, but due to the crash neutral. And this crash was more severe than the one in the empty roundabout.\\
 P29 & Unsafe &  Car didn't respect the stops on its own\\
 P30 & Neutral &  I had to interract to many times becaus I did not feel so safe\\
 P31 & Neutral &  Except during the situation whe the car in front behaved strangely (which caused an accident), i felt save the whole time. Since I did not have control in previous situations, I already trusted the AI.\\
 P32 & Unsafe &  Better decisions needed when other cars are around,\\
 P33 & Unsafe &  The car changed lane in middle of the road in the start. It went above the roundabout at the end.\\
 P34 & Neutral &  Better, but in the end it still hit the curb\\
 P35 & Neutral &  I was again able to slow down the car. But it still didnt manage to drive the roundabout.\\
 P36 & Safe &  i could controll the speed, so less suddent breaks with the slow car in front\\
 P37 & Safe &  It was safe because I could intervene when the car was making mistakes and I could avoid the usual collision at the round about. \\
 P38 & Very Safe &  Speed Influence enhances safety feeling\\
 P39 & Safe &  safe and easy to use\\
 P40 & Very Unsafe &  Two matters: 1) driver keeps its distance to the can in the front, but with sharp breaks instead of slowing down the car. 2) inable to avoid strange behaviors and drove next to a car with unstable drive and had an accident.\\
 P41 & Neutral &  no irritation because of other obstacles or weather,  but still unsteady/unsafe behaviour in the end, lots of intervention from me\\
 P42 & Very Unsafe &  Caused a collision\\
 P43 & Unsafe &  Crashed in roundabout and drove with too much speed in it, ignored stop signal\\
 P44 & Neutral &  slowing down before the STOP, moderating the speed when the rain starts and when entering the roundabout \\
 P45 & Very Unsafe &  car doesnt drive safely without adjusting the max speed yourself\\
 P46 & Safe &  i was able to control the speed and thus avoid a person on the bicycle. this gave me back some control.\\
 P47 & Very Unsafe &  -\\
 P48 & Safe &  having the control to react to different situations feels safe\\
 P49 & Safe &  regulating the speed of the car gives a strong sense of safety because i can keep distance from an unpredictable driver, this way i was not so much disturbed by the bad driving and just slowly followed. after that i could speed up again and slow down again at the roundabout.\\
 P50 & Very Unsafe &  The car left the road and drove over the sidewalk. Afterwards, it crashed into a streetlight. It seemed like it wanted to avoid the other car that had crashed, but overcorrected and left the road.\\
    \bottomrule
    \end{tabular}}
\end{table}

\clearpage
\subsection{\RQ{2}: Without interaction}
\begin{table}[h!]
    \centering
    \scriptsize
    \caption{Non-interactive scenario without obstacles}
    \resizebox{0.8\linewidth}{!}{
    \begin{tabular}{llp{10cm}}
    \toprule
    \textbf{Code} & \textbf{Perceived Safety} & \textbf{[C3] Justify the perceived safety of  scenario 1?} \\ \midrule
 P1  & Unsafe & the car, as mentioned before has to fast steering actions time to time. Going in vertical path is safe in general, mainly at the end the car bumped on the merging of the road this made this scenario unsafe \\
 P2  & Neutral & The car drove more moderate but the turns were a bit too sharp. \\
 P3  & Very Unsafe & the car took strange ways and it crashed in the end \\
 P4  & Neutral & It felt safe all the path except the roundabout where the car went off-road. \\
 P5  & Unsafe & it was too fast in the circle \\
 P6  & Unsafe & unsafe \\
 P7  & Unsafe & Vehicle struck the side area of roundabout. Didn't lookedsafe  \\
 P8  & Neutral & Safe in the beginning, but again climbed the roundabout at the end. The sound of glass breaking when the car mount the roundabout is a nice touch. \\
 P9  & Unsafe & am Ende des Tests: Glass ging zu Bruch \\
 P10 & Unsafe & it follows traffic light but got distracted on square and not able to follow path \\
 P11 & Unsafe & the car ran on the foot path \\
 P12 & Unsafe & Followed rule but have issue at roundabout. \\
 P13 & Neutral & Was good until it went on the curb aat the end \\
 P14 & Safe & Overall a smooth driving experience except on the round about \\
 P15 & Neutral & Soft crash in the end, everything was good until then. \\
 P16 & Neutral & not smooth with obstacles at a very few  moments.  \\
 P17 & Unsafe & The car only stops at some red light and ignores some stop markings on the road. \\
 P18 & Safe & The car did a good job of staying within the limits and following the traffic rules \\
 P19 & Unsafe & Overall turning was good but again as in the previous two scenarios, the car would hit the sideways while moving left in the roundabout \\
 P20 & Neutral & in the end it hits the side of the roundabout \\
 P21 & Unsafe & He didnt stop at the STOP sign and at the end I heard the sound of crashing behind \\
 P22 & Unsafe & It felt pretty safe until something crashed on the back of the car, it also stopped in the middle of the road after it \\
 P23 & Very Unsafe & The car did not stop at a stop marking, it also had problems at the randabout. \\
 P24 & Unsafe & ran over stop sign; hit the curb \\
 P25 & Neutral & The car was keeping a low speed so it felt quite ok the overall experience even if it had problem with the roundabout were it hits the obstacle with the left side of the car. The feeling of being in the car it feels quite real. \\
 P26 & Unsafe & The car hitted at the end.  \\
 P27 & Unsafe & Although the car seemed to follow the traffic rules and speed limits, it was not able to navigate the round-about, crashed on to the curb and lost control in the end. \\
 P28 & Very Safe & driving extremely slow through an empty city, nothing to be worried about. \\
 P29 & Very Unsafe & Car didn't respect the stops and and been damaged \\
 P30 & Safe & It was not so bad even when the car hit the border in the circle \\
 P31 & Very Safe & From within the car, it felt more realistic. The car did keep on track well and stopped at the road crossings and signals. Since it did not drive too fast, it felt really save.  \\
 P32 & Safe & Moderate. Safelz driven. \\
 P33 & Very Unsafe & The car changed lane in middle of the road in the start. It  hit the roundabout at the end and got into an accident (based on sound experience but could not see anything). \\
 P34 & Unsafe & drove on curb, too slow, ignored stop \\
 P35 & Neutral & No slowing down for the intersection and later crashed in the roundabout \\
 P36 & Unsafe & this time i was more concerned about the first turn, car did not really stop at the intersection to look for other cars crossing. also the hit to the round about in the last second, felt reallt unsafe.  \\
 P37 & Unsafe & Driving is smooth, but I took the sidewalk and apparently something broke.  \\
 P38 & Very Safe & Altought the steering is not very smoothly, it was very relaxing to drive \\
 P39 & Safe & it was smooth experience \\
 P40 & Neutral & I heard a crash towards end of the experiment, but I felt safe since there was not only other objects around. \\
 P41 & Neutral & most of the time correct in line behaviour, but very slow feeling (maybe due to latency of simulation), touching of roundabout in the end made it somewhat unsafe again \\
 P42 & Very Safe & Weird roundabout interaction, otherwise very safe. \\
 P43 & Very Unsafe & Car crashed and ignored stop sign. \\
 P44 & Unsafe & skipping the STOP sigh and invading the vegetation in the roundabout  \\
 P45 & Unsafe & At the end the car bumped into the roundabout. \\
 P46 & Unsafe & the car did not stop on stop sign. the car crashed in the roundabout. this made me feel unsafe.  \\
 P47 & Very Unsafe & - \\
 P48 & Neutral & safe until the end, but no other traffic members are involved  \\
 P49 & Safe & waas quite okay, respects all the traffic rules. just the roundabout is a bit stressful and would have liked to be able to control the steering wheel \\
 P50 & Very Safe & The car drove correctly and followed the rules. \\
    \bottomrule
    \end{tabular}}
\end{table}

\clearpage
\begin{table}[h!]
    \centering
    \scriptsize
    \caption{Non-interactive scenario with obstacles}
    \resizebox{0.8\linewidth}{!}{
    \begin{tabular}{llp{10cm}}
    \toprule
    \textbf{Code} & \textbf{Perceived Safety} & \textbf{[C3] Justify the perceived safety of 2?} \\ \midrule
 P1  & Unsafe & similar level of unsafety. main difference was that the not so smooth behavior of the car happens also in the proximity of other cars (fast restarts followed by rapid stops, a more safe driving would be ideal). However, without obstacles i felt in some cases the car was too slow compared to the one of BEAMNG \\
 P2  & Unsafe & The car should drive more conscious in the dark. \\
 P3  & Very Unsafe & the car drove really strange, it sped up and down and did not drive consistently \\
 P4  & Neutral & It felt very safe all the way, except the roundabout where it went off-road. The fact that it stops when there is a motorbike in front and that it obeys traffic lights makes it increase the safety perceived. \\
 P5  & Unsafe & it was too fast in the circle \\
 P6  & Neutral & not very safe \\
 P7  & Very Unsafe & It was very unnsafe. The vehicle was not able to take turn on the roundabout and there was accident. \\
 P8  & Neutral & Good in the beginning, the turn when the weather was dark was a bit unnatural, climbed the pavement at the end \\
 P9  & Unsafe & am Ende des Tests: Glass ging zu Bruch, es regnete IM Auto, zu nahe auf Feuerwehrfahrzeug aufgefahren \\
 P10 & Very Unsafe & car crashed in another car when other car does not follow usual path and after crash not able to identify the traffic light \\
 P11 & Neutral & the car complted the track  \\
 P12 & Very Unsafe & Cannot handle sudden change of situation. \\
 P13 & Very Unsafe & didnt properlz assess the vehicle in front and didnt brake when the vehicle came in front of it not did it stop \\
 P14 & Safe & Decent work despite challenging obstacles \\
 P15 & Neutral & Soft crash in the end, everything was good until then. Also, close call with another car. \\
 P16 & Unsafe & Did not hit any other car/ person but at a round about it hit the ramp and ended in a bush.  \\
 P17 & Unsafe & The car ignores stop markings on the street and veeres off the road. \\
 P18 & Neutral & The did a good job of being within the speed limits but lost control at a turn and crashed the rear end \\
 P19 & Unsafe & for the majoity of the ride, the drive was safe. It was responding well to the abnormal behaviour of bike applying brakes abruptly. Towards the end, it again hit the sideways at the roundabout. This made it unsafe.  \\
 P20 & Neutral & it had an accident but it was not its fault \\
 P21 & Very Unsafe & Because he didnt slow down smothly for the bycicle and at teh end the crashing sound was uncomfortable \\
 P22 & Unsafe & Same as in scenario 1, it was safe at the beginning with the other cars also but at the end it crashes and it stops in the middle of the road  \\
 P23 & Unsafe & The car accelerates and stops abruptly. I know it is because the care in front was doing exactly the same. In such situations, I would simply drive at a safer distance from the driver in front of me. \\
 P24 & Very Unsafe & instantly crashed into curb and lamp post \\
 P25 & Very Unsafe & The car was breaking and accelerating a lot while being behind the other car and also the other car was not behaving safely on the road, ending the simualtion with an accident between the two, so it felt quite unsafe overall.  \\
 P26 & Very Unsafe & The distance betweent the car and the obstacles was very close.  \\
 P27 & Unsafe & It was safe for the most part, but the car lost control in the round-about and crashed on to the curb. \\
 P28 & Very Unsafe & it drove extremely close up to the ambulance car and finally crashing into it. therefore, the worst case happen. \\
 P29 & Very Unsafe & Car didn't stop at stops and was outside the lines in the curves, brakes were very abrupt \\
 P30 & Neutral & Turning around was to fast. I was still looking to the left if there is a car coming and looking out infront of the car it was a big surpise seeing alt he cars infront of me. Then we had a nice accident xD \\
 P31 & Neutral & From within the car, it felt more realistic. The car did keep on track well and stopped at the road crossings and signals. Since it did not drive too fast, it felt really save.  \\
 P32 & Very Unsafe & Car overturned. \\
 P33 & Very Unsafe & The car changed lane in middle of the road in the start. It  hit the roundabout at the end and got into an accident (based on sound experience but could not see anything). \\
 P34 & Unsafe & drove on curb, too slow, ignored stop \\
 P35 & Neutral & No slowing before the intersection and crash in the roundabout. But did stop for red light. \\
 P36 & Safe & safe, but still it could handle the slow dirver in front better, too much use of hard break.  \\
 P37 & Unsafe & "I only feel unsafe because something broke again at the end. I do not feel unsafe that it stops all the time because of the motorcycle, but the guy on the motorcycle obviously does not know how to drive and the car should double cross him. "  \\
 P38 & Very Safe & The only thing was, that the car moves not precise enought \\
 P39 & Very Unsafe & taking turn was poor and car crashed \\
 P40 & Unsafe & There was a motobycle with very strange behavior which was ignored by the drive until the accident happened. \\
 P41 & Very Unsafe & started okay, but the  unsteady behaviour of other vehicles made me feel uncomfortable, late stopping of ego vehicle also, very unsafe behaviour around roundabout \\
 P42 & Very Safe & Except the roundabout issue, very safe. \\
 P43 & Very Unsafe & Ignored stop sign, had a crash into roundabout.  \\
 P44 & Very Unsafe & not smooth STOP and GO, entering the lane with other road users in unsafe manner, \\
 P45 & Unsafe & Car drove into roundabout at the end. \\
 P46 & Unsafe & the car did crash with an ambulance and did not stop on stop sign. in the end it crashed in the round about.  \\
 P47 & Very Unsafe & - \\
 P48 & Neutral & rather safe, no reaction to other cars though \\
 P49 & Unsafe & the car crashing felt unsafe and i would have liked that the car would make a bigger curve around it. and then crashing in the house was not great either \\
 P50 & Safe & The car drove correctly, however, at one point, it was moving wobbly, which it immediately course corrected. \\
    \bottomrule
    \end{tabular}}
\end{table}

\clearpage
\section{\RQ{3}: Comments on realism}

\subsection{Realism for \beamng}
\begin{table}[h!]
\centering
\scriptsize
\caption{Comments on realism of the test cases in \beamng}
\resizebox{12.5cm}{!}{
\begin{tabular}{llp{12cm}}
\toprule
\multirow{2}{*}{\textbf{Code}} & \textbf{Perceived} & \multirow{2}{*}{\textbf{Justify level of realism of scenarios generated test cases?}} \\
                & \textbf{realism} &  \\\midrule
P1   & 3 & ``With the road including obstacles the realist of the scenario was ok. The level of realism of the experience drastically improved when I am within the car. perceived level of safety reduced drastically with the increase level of realism.''   \\
P2   & 4 & ``The simulation was smooth and there was a high resolution.'' \\
P3   & 2 & ``the grass the horizon as well as the red vertical lines do not very realistic'' \\
P4   & 4 & ``The realism is quite good, especially in the car design. The car structure was damaged after crashing, the wheels were getting broken, and there was smoke coming out. The inside view of the car was also prety real, with the driver's hand moving the steering wheel, and all the car panel commands. The reason I don't put it a 5 is that the landscape and road design was not as real, especially when compared to some videogames such as GTA5 or red-dead redemption.'' \\
P5   & 2 & ``Driving was too careless'' \\
P6   & 4 & ``normal'' \\
P7   & 4 & ``Being in simulator, I can see that simulator took care of all the aspects of real life scenario like when it crashed carswerved. Inside the car view was also very as per driver reference and it felt safe and like how driver will drive'' \\
P8   & 4 & ``It was not too realistic, but it was also not too shabby. The roads specifically felt very real, but the environment itself did not feel too polished.'' \\
P9   & 4 & ``good genough for this test'' \\
P10  & 3 & ``without obstacles path felt more real. Obstacles realism can be improved'' \\
P11  & 4 & ``the test cases has obsticals and real word road conditions'' \\
P12  & 2 & ``They didnt have any difference in scenarios.'' \\
P13  & 2 & ``the roads were as thez would be in actual life.'' \\
P14  & 4 & ``The driver seat simulator felt very realistic'' \\
P15  & 5 & ``Good real-life simulation'' \\
P16  & 4 & ``with empty road, it felt very real.'' \\
P17  & 3 & ``the car seems to behave realisticly, but the environment isn't very real life-like yet.'' \\
P18  & 4 & ``  It replicates real world scenarios quite well'' \\
P19  & 4 & ``the view was realistic and not distorted.'' \\
P20  & 4 & ``the crashes are more realistic, a physical object got stuck between the wheel and the engine for example.'' \\
P21  & 4 & ``it was diffrent when I sit in the car than from outside so it felt more real. But still looked like a game so not that realistic.'' \\
P22  & 4 & ``They respect the scale from the objects'' \\
P23  & 4 & ``I think that the roads generated and the obstacles were pretty realistic.'' \\
P24  & 3 & ``The car didnt bechave as expected when driving on grass, other than that it was quite realistic'' \\
P25  & 3 & ``The scenario felt ok, especially the one with the obstacles and the bumps of the car on the obstacles on the road. Same for the quality of the perception in the simulation inside the car. It would be nice to add the sound to the scenario to increase realism.'' \\
P26  & 2 & ``The obstacles are static. Also, the way seems unrealistic.'' \\
P27  & 4 & ``The movement of the car, speeding up and down, reaction to going over bumps etc. seemed very realistic.'' \\
P28  & 1 & ``It looks like a computer game to me and I can differentiate clearly between reality and fiction as far as I know.'' \\
P29  & 4 & ``Very simple scene, few details'' \\
P30  & 2 & ``The steets were not so realistic'' \\
P31  & 4 & ``The overall level of realism regarding the behaviour of the car is quite good from the perspective inside the car. The surroundings in the environment do not look very realistic compared to state-of-the-art computer games that focus on car driving.'' \\
P32  & 3 & ``good'' \\
P33  & 4 & ``Someitme it is hard to know the expected outcome.'' \\
P34  & 5 & ``good length of the curves and good obsticals, that would be a challange for humans as well'' \\
P35  & 3 & ``The environment with the obstacles had way to many speed bumps for it to be realistic. The other environment was ok.'' \\
P36  & 2 & ``the road was not realisitic in some cases, like a very small side road in warmup scenario.  speed bumps are too close to each other, and the obstacles in the middle feel artificial.'' \\
P37  & 2 & ``There is nothing in the environment, it feels like a test route for car industry.'' \\
P38  & 4 & ``In real the steering is much more precise'' \\
P39  & 4 & ``It seems good enough and can feel the environment'' \\
P40  & 3 & ``Low graphics from the environment, no other cars or people around, car feels quite regid and going over the obsticles is not bumpy.'' \\
P41  & 3 & ``testcase without obstacles seems more realistic, second one has quite a lot bumps, but might be realistic for some roads'' \\
P42  & 4 & ``Real life scenarios with bumpy or uneven roads and obstacles.'' \\
P43  & 3 & ``Movement is not as smooth as one would usually associate with a car, it moves quite jerkily. The general feel of the simulation is slightly animated and not as real looking'' \\
P44  & 4 & ``good visibility of the lines and realism of the speed'' \\
P45  & 3 & ``Looks like a game, not the real world.'' \\
P46  & 3 & ``there ar no other cars on the street. basic surroundings.'' \\
P47  & 3 & ``Bin ??berfragt.'' \\
P48  & 3 & ``realistic scenarios'' \\
P49  & 4 & ``when i was sitting in the drivers seat i was actually quite uncomfortable and it did feel like it was real, except that my body of course does not move similarly. but i noticed that i was physically reacting to the things happening on the screen.'' \\
P50  & 4 & ``It is realistic to have curvy roads, and the obstacles simulate other cars decently enough for testing whether the car would be a hazard for other cars.'' \\
\bottomrule 
\end{tabular}}
\end{table}
\clearpage
\subsection{Realism for \carla}

\begin{table}[h!]
\centering
\scriptsize
\caption{Comments on realism of the test cases in \carla}
\resizebox{12.5cm}{!}{
\begin{tabular}{llp{12cm}}
\toprule
\multirow{2}{*}{\textbf{Code}} & \textbf{Perceived} & \multirow{2}{*}{\textbf{Justify level of realism of scenarios generated test cases?}} \\
                                & \textbf{realism}                  &       \\ \midrule
P1   & 5 & It was possible to observe almost anything you see in a city (pedestrian, cars, other vehicles, traffic signs, etc.). It was also way more realistic driving style. \\
P2   & 2 & The simulation was very haltingly and it felt in general very artificial. The sound of the car was just weird, especially for an electrical car. \\
P3   & 4 & it had sounds and a lot of world objects \\
P4   & 4 & Very good actually. I don't put a 5 because there is always room for improvement and I've seen game engines with more realistic results, but I was positively surprised. While the car was designed as a single box, the landscape was much more realistic, which made you more immerse in the scenario. Also the fact that it uses full VR (3D) makes a big difference. \\
P5   & 4 & it was good but didnt stop on the stop signals \\
P6   & 5 & good \\
P7   & 5 & The scenario was very real along with traffic lights, day and night, foggy and it looked like high quality graphics. \\
P8   & 5 & This level was more realistic compared to the previous simulator.  \\
P9   & 4 & goo for this test \\
P10  & 4 & scenarios seems very real \\
P11  & 3 & it covered the scenarions like trafic light, city limit  \\
P12  & 5 & AI was very responsive \\
P13  & 5 & the world and the objects in it were very realistic including the sounds \\
P14  & 4 & It gave a clear sense of traffic rules and turns although you can expect much more traffic and dangers in the real world. \\
P15  & 5 & It was very realistic \\
P16  & 5 & The city graphics look very real.  \\
P17  & 4 & The environment (lighting, obstacles) feel quite real. \\
P18  & 5 & Replicates the real world scenario very well. \\
P19  & 3 & reality seemed okay  \\
P20  & 4 & best than BeamNG t generate realistic environments and scenarios, as well as a better graphical aspect \\
P21  & 5 & Compared to the first simulation it was better and more realistic because of the enviroment. \\
P22  & 5 & The scale of the objects and details are really good \\
P23  & 5 & It was very realistic, because it was in a city, there were other vehicles and the weather was changing. \\
P24  & 1 & The movements were very abrupt, the physics did not feel too realistic \\
P25  & 5 & The scenario it quite realistic probably also because it consider the simulation inside a city in which the car has to deal with traffic signs and other cars, but also people. \\
P26  & 3 & Better for shadows, textures, and realistic obstacles. However, phisics still not realistic enough. \\
P27  & 3 & There were some lags in the animation and the movement of the car did not seem realistic at times. During the rainy weather scenarios, raindrops were falling inside the car even though everything was closed. \\
P28  & 3 & still perceivable as a computer simulation, but with much more details and therefore kind of comparable to reality. \\
P29  & 4 & Scene isn't too detailed \\
P30  & 4 & It was a good scenario expect some cars which where driving in a snale line. \\
P31  & 5 & The surroundings have more detail which made it feel more realistic. \\
P32  & 4 & OK \\
P33  & 4 & Changing lane was not safe \\
P34  & 5 & Very nice with the real city map \\
P35  & 4 & It was more realistic than BeamNG since it was an actual city. The car was also driving smoother which helped for the realism. \\
P36  & 4 & the city environment and the obstacle, cars, etc. were realitic. just the drivers of bicycle and cars were too artifically crazy in some cases. you don't expect that level of crazyness on the real streets normally \\
P37  & 5 & there are cars and environment on the side and there are street lights and stop signs. Too bad there are no bytanders \\
P38  & 4 & They were very interactive and quite realistic \\
P39  & 4 & scenes are really good \\
P40  & 4 & The simulation is quite realistic including sound. The resolution of traffic signs and lights (far objects) are too low and direction lines on trafic lights and sometime on the streets are missing. \\
P41  & 4 & it has some latency, which gives an akward feeling, but was better than the other simulator \\
P42  & 5 & Great generation and simulation \\
P43  & 3 & Cars do not follow all traffic rules. They ignore stop signs and do not indicate when to turn. But the being driven feels realistic to me, just like being a passenger. \\
P44  & 4 & well reproduced city conditions (infrastructure, buildings, vegetation, etc). Not always well understood behaviours of other vehicles, weather conditions included \\
P45  & 4 & Quite a lot of details but you can see it is obviously not real world. \\
P46  & 4 & the buildings, different cars / bicycles / persons. the street marking etc.  \\
P47  & 3 & Umgebung war ziemlich realistisch. Jedoch fehlt das grosse Verkehrsaufkommen. \\
P48  & 4 & very realistic environment \\
P49  & 5 & felt real, because it is really what a city or village could look like so i really reacted to what was happening on screen. \\
P50  & 5 & The scenarios were very realistic, complete with other vehicles that weren't driving correctly all the time (which is accurate to real life). \\
\bottomrule 
\end{tabular}}
\end{table}